\documentclass{aa}  

\usepackage{graphicx}
\usepackage{txfonts}
\newcommand{\re}{$R_e$}
\newcommand{\Lsig}{$L-\sigma$}

\newcommand{\Lsigb}{$L=L'_0\sigma^{\beta}$}
\newcommand{\Lsigbtempo}{$L=L'_{0}(t) \sigma^{\beta(t)}$}                   

\newcommand{\MR}{$\log(R_e)-\log(M_s)$}
\newcommand{\MRa}{$R_e$--$M_s$}
\newcommand{\Rsigma}{$R_e - \sigma$}

\newcommand{\muem}{$\langle\mu\rangle_e$}

\newcommand{\Ie}{$I_e$}

\newcommand{\IeRe}{$I_e - R_e$}
\newcommand{\IeSig}{$I_e - \sigma$}

\newcommand{\FPR}{$\log(\sigma)-\log(I_e)-\log(R_e)$}

\newcommand{\ie}{{i.e.}}
\newcommand{\eg}{{e.g.}}

\begin{document}

   \title{A new framework for understanding the evolution \\ 
   of early type galaxies}


   \author{M. D'Onofrio\fnmsep\thanks{Corresponding author: Mauro D'Onofrio}
          \inst{1}
          \and
          C. Chiosi\inst{1}
          }

   \institute{Department of Physics and Astronomy, University of Padua,
              Vicolo Osservatorio 3, I35122 Padova (Italy)\\
              \email{mauro.donofrio@unipd.it} \\
              \email{cesare.chiosi@unipd.it}
             }

   \date{Received December, 2022; accepted January, 2023}

 
  \abstract
{We have recently suggested that the combination of the scalar virial theorem ($M_s \propto R_e \sigma^2$) and the \Lsigb\ law, with $L'_0$ and $\beta$ changing from galaxy to galaxy (and with time), can provide a new set of equations valid for investigating the evolution of early-type galaxies \citep{Donofrio_Chiosi_2022}. These equations are able to account for the tilt of the Fundamental Plane and to explain the observed distributions of early-type galaxies in all its projections.}
{In this paper we analyze the advantages offered by those equations, derive the $\beta$ and $L'_0$ parameters for real and simulated galaxies, and demonstrate that, according to the value of $\beta$, galaxies can move only along some permitted directions in the fundamental plane projections. Then, we show that simple galaxy models that grow in mass by infall of gas and form stars with a star formation rate depending on the stellar velocity dispersion nicely reproduce the observed distributions of early-type galaxies in the Fundamental Plane projections and yield $\beta$s that agree with the measured ones.}
{We derive the mutual relationships among the stellar mass, effective radius, velocity dispersion, and luminosity of early-type galaxies as a function of $\beta$ and calculate the coefficients of the Fundamental Plane. Then, using the simple infall models, we show that the star formation history of early-type galaxies is compatible with  the  $\sigma$-dependent star formation rate, and that both positive and negative values of $\beta$ are possible in a standard theory of galaxy evolution.}
{The parameter $\beta(t)$ offers a new view of the evolution of early-type galaxies. In brief, i) it gives a coherent interpretation of the Fundamental Plane and of the motions of galaxies in its projections; ii) it is the fingerprint of their evolution; iii) it measures the degree of virialization of early-type galaxies; iv) and finally it allows us to infer their evolution in the near past. }
   {}
   

   \maketitle
%

\section{Introduction}
\label{sec:1}

This study is the latest of a series aimed at demonstrating that the scaling relations (Sc-Rs) for early-type galaxies (ETGs), \ie\ the mutual correlations between the main structural parameters of galaxies (\eg\ the stellar mass $M_s$, the effective radius \re, the effective surface intensity \Ie, the luminosity $L$ and the central velocity dispersion $\sigma$)\footnote{Therein after by structural parameters of a galaxy we mean those of the above list, and leave aside the parameters that define the internal structure such as the S\'ersic index, the axial ratio, etc. Galaxies are considered point mass objects.}, can be fully understood if we adopt a new perspective in which  the Virial Theorem (VT) of the stellar systems is coupled to the galaxy luminosity taking into account that this latter can randomly vary with time as a result of accretion/depletion events associated to mergers/close encounters  expected in the hierarchical galaxy  formation scenario in addition to the natural evolution of its stellar content.  The new equation governing the luminosity is expressed by
\begin{equation}
 L(t) = L'_0(t) \sigma(t)^\beta(t).
 \label{eq1}
\end{equation}
in which $L$, $L'_0$, $\sigma$ and $\beta$ are all functions of time and can vary from galaxy to galaxy. 
This relation is formally equivalent to the Faber \& Jackson relation for ETGs \citep{FaberJackson1976}, but it has a profoundly different physical meaning. In this relation $\beta$ and $L'_0$ are free time-dependent parameters that can vary considerably from galaxy to galaxy, according to the mass assembly history and stellar evolution of each object. This relation empirically encrypts  the  effects of all the above physical processes  in terms of luminosity and velocity dispersion variations, parameters that can both vary across time because galaxies evolve, merge, and interact. 

In our previous works we tried to highlight some of the advantages offered by coupling the  VT with the \Lsigb\ law. The first efforts were dedicated to understand the origin of the Fundamental Plane (FP) of ETGs and the distributions observed in its 2D projections \citep{donofrioetal17,Donofrioetal2019,Donofrioetal2020,DonofrioChiosi2021}. 
While discussing this problems,
\cite{Donofrioetal2017} and \cite{Donofrio_Chiosi_2022} advanced the idea that the explanation invoked for  the origin of the FP tilt (and its small scatter) should also account for the observed distributions of galaxies in all the 2D projections of the FP. The solution was found in the coupling of the VT with the time-dependent \Lsigb\ relation. 

The key idea behind this approach is that the luminosity of galaxies is not simply related to the total stellar mass, but also to random variations caused by mergers and interactions. This implies that, accepting the \Lsigb\ law as an empirical descriptor of the possible changes occurring in $\sigma$ and $L$, one can describe a galaxy with two different independent equations: the classical scalar VT on the notion  that galaxies are always very close to the mechanical equilibrium, and the \Lsigb\ law  that  fully accounts for all possible processes taking place during the lifetime of a galaxy. 

Our previous studies have successfully shown that a $\beta$ parameter changing  with time and assuming either positive and negative values, can easily explain the movements and distribution in the planes of the Sc-Rs. This approach is in fact able to explain in a natural way the tilt of the FP, the existence of the Zone of Exclusions (ZoE) observed in many Sc-Rs, and the direction of motion derived from the changes in $\sigma$ and $L$.             

In this work we aim to provide evidences that such an approach gives a global interpretation of the Sc-Rs observed for ETGs and that even the classical monolithic view of mass assembly is in agreement with the idea of a variable $\beta$ parameter thus  confirming the  \Lsigbtempo\ law. 

The paper is organized as follows: Sec. \ref{sec:2} gives a short description of the samples of galaxies (both real and simulated) used in this work; Sec. \ref{sec:3} is dedicated to the derivation of the new equations of galaxy evolution and to the different relations among the structural parameters in all FP projections; Sec. \ref{sec:4} presents a few new simple models of ETGs growing with a SFR depending on $\sigma$ and accounts for the role of $\beta$. Finally, Sec. \ref{sec:5} provides our discussion and conclusions. In all calculations we used the parameters of the $\Lambda$CDM cosmology.

\section{The samples of real and model galaxies}
\label{sec:2}

\textsf{ Observational data}.
The observational data used in this work are the same of  \cite{DonofrioChiosi2021} and \cite{Donofrio_Chiosi_2022}. The data for the real galaxies are extracted from the WINGS and Omega-WINGS databases
\citep{Fasanoetal2006,Varela2009,Cava2009,Valentinuzzi2009,Moretti2014,Donofrio2014,Gullieuszik2015,Morettietal2017,Cariddietal2018,Bivianoetal2017}. 

The sample is not homogeneous because the spectroscopic database is only a sub-sample of the whole optical sample. The ETGs with available velocity dispersion $\sigma$, stellar mass $M_s$, and star formation rate SFR, are less numerous than those extracted from the photometric database (providing \re, \Ie, $n$, $L_V$, etc.). 

In particular we used:
1) the velocity dispersion $\sigma$ of $\sim1700$ ETGs. The $\sigma$ {measurements come from the SDSS and NFPS databases \citep{Bernardietal2003,Smithetal2004}} and were measured within a circular area of 3 arcsec around the center of the galaxies;
2) the luminosity, effective radius and effective surface brightness in the V-band of several thousand ETGs, derived by \cite{Donofrio2014} with the software GASPHOT \citep{Pignatelli}. The effective radius is determined from the luminosity growth curve by considering the circle that contains half the total luminosity. The effective surface intensity follows directly from the knowledge of $L$ and \re;
3) the distance of the galaxies derived from the redshift measured by \cite{Cava2009} and \cite{Morettietal2017};
4) the stellar mass obtained by \citet{Fritzetal2007}, only for the galaxies of the southern hemisphere. 

The cross-match between the spectroscopic and optical samples provides here only 480 ETGs with available stellar mass, luminosity, velocity dispersion, S\'ersic index, effective radius and effective surface brightness.
The error of these parameters is $\simeq20\%$. These are not shown in our plots, because they are much lower than the observed range of variation of the structural parameters in the scaling relations and do not affect the whole distribution of ETGs.

Occasionally, we have also used the catalog by \citet{Burstein_etal_1997} containing objects from Globular Clusters (GCs), Dwarf Galaxies (DGs) of different types, to late and early type galaxies (LTGs and ETGs, respectively), and finally clusters of galaxies (GCGs). They are used to have a general idea of the Sc-Rs for systems of different sizes, but dynamically close to the virial condition. Limited to ETGs sometime we also used the sample of \citet{Bernardi_etal_2010}. 

\textsf{Simulated galaxies}. The hydrodynamic simulations, are probably the best galaxy models today available to compare theory with observations despite the fact that several problems still bias their results. There are several suites of galaxy simulations in cosmological context among which we  recall Illustris-1 by \citet[][]{Vogelsbergeretal2014,Genel_etal_2014,Nelsonetal2015},  recently superseded by Illustris-TNG by 
\citet{Springeletal2018,Nelson_etal_2018,Pillepich_etal_2018a}, and EAGLE by \citet{Schaye_etal_2015}.  We decide to adopt here Illustris-1 for two reasons: first chief the fact we want to be consistent with the results shown in our previous papers on this same subject that were based on the Illustris-1 models. Second, we have checked that the main results of our analysis do not change passing from Illustris-1 to Illustris-TNG. 

The kind of analysis carried out here is indeed somehow  independent of the level of precision reached by models from different sources, because we are mainly interested to present a new method for deciphering the information encrypted in the observational data about the past history of ETGs.  To this aim, we have extracted from the Illustris-TNG database at redshift $z=0$ a sample of about thousand model galaxies of all possible masses that are used to support the above statement.  

Our data-set extracted from Illustris-1 consists of several sub-sets of about $\sim 2400$ galaxies each,  sampled different at redshifts from to $z=0$ to $z=4$. A full description of these data is given in \cite{Cariddietal2018} and \cite{Donofrioetal2020}. In particular, we collected  the effective radii, the total luminosity, the stellar mass, and the velocity dispersion, the age, and the star formation rate, together with radii, masses, and velocity dispersion of the dark matter component.

  \begin{figure*}
   \centering
 {  \includegraphics[scale=0.45]{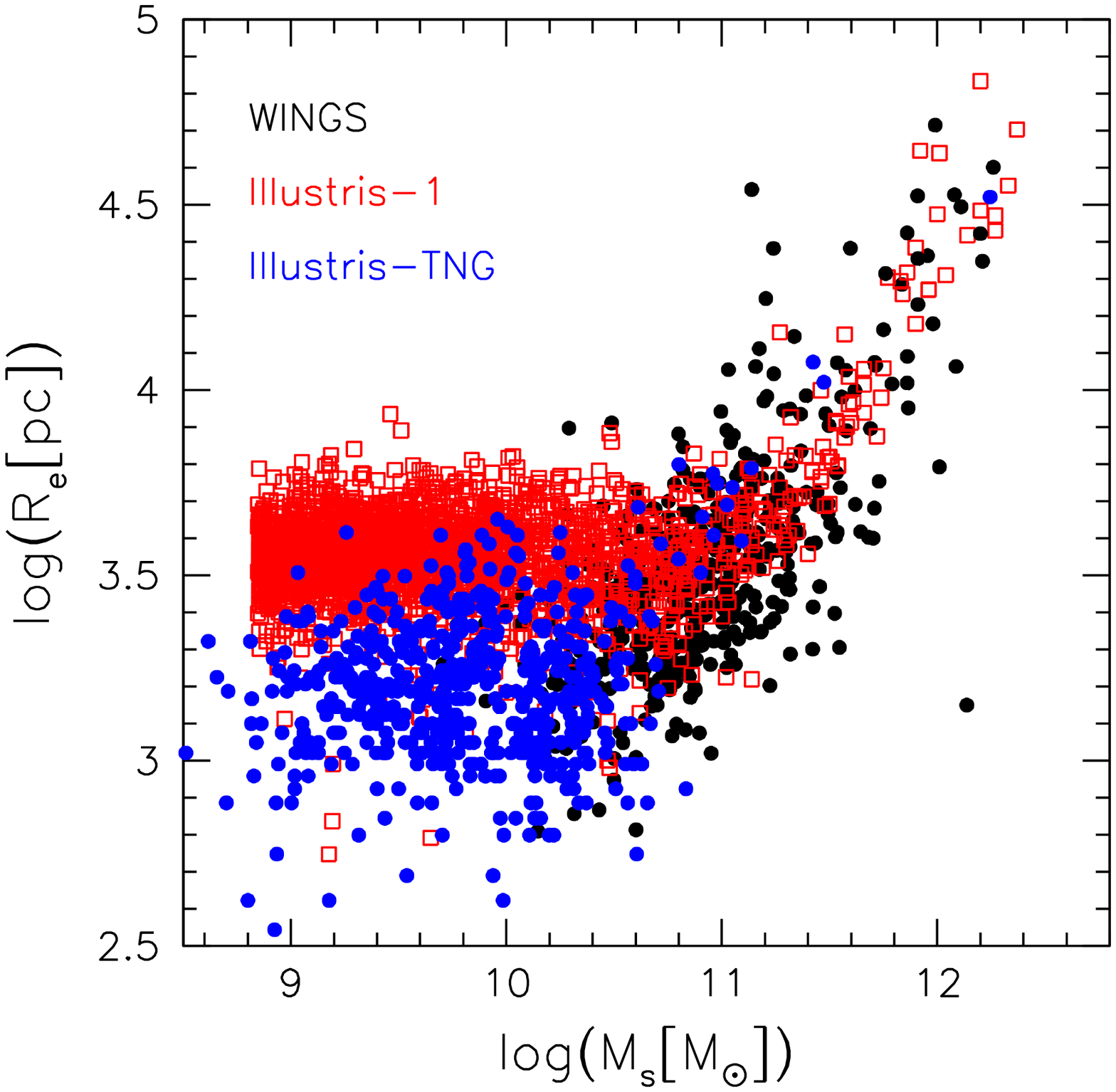}
   \includegraphics[scale=0.45]{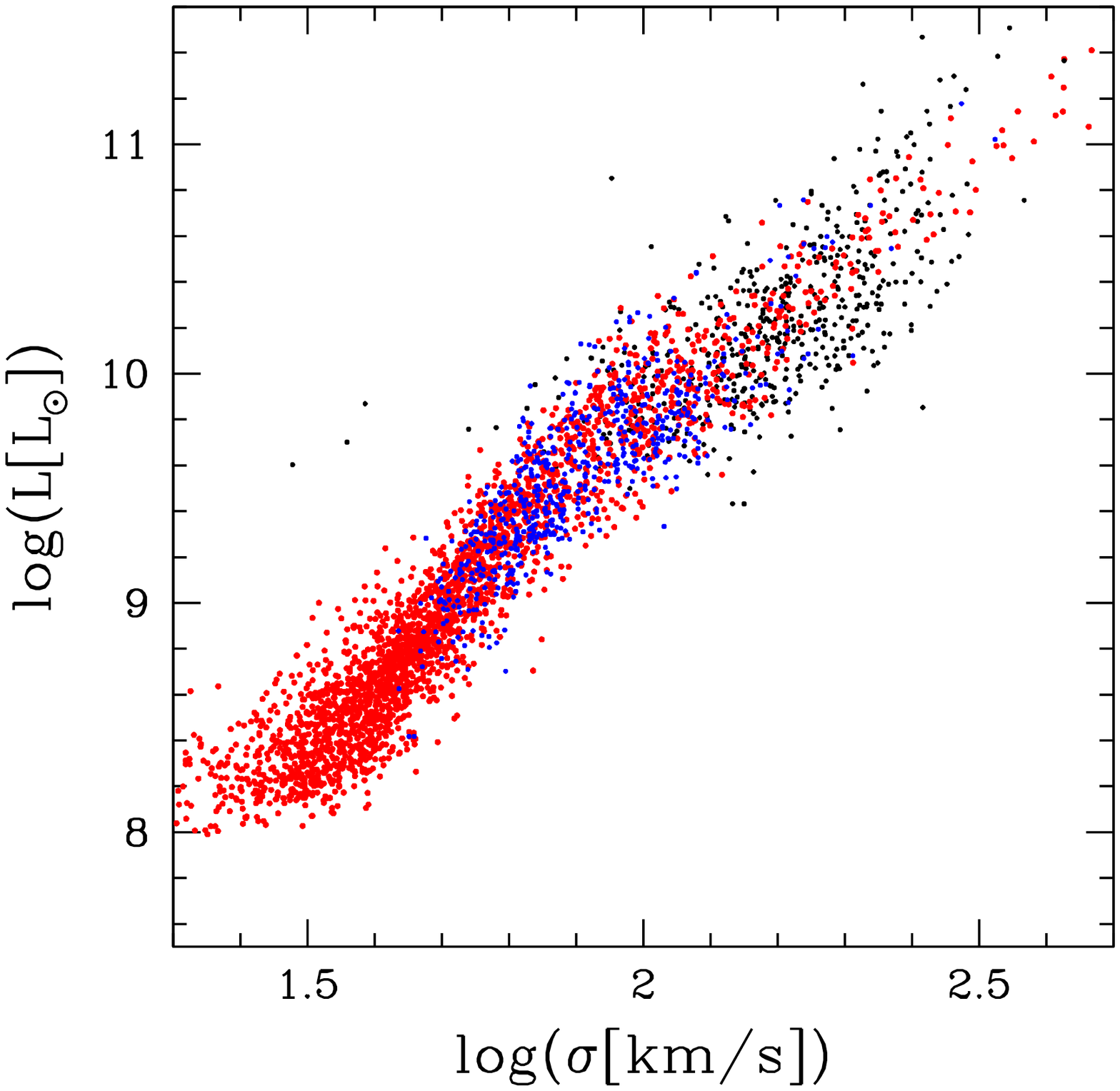} }
   \caption{Left panel: The stellar mass versus radius relations for the Illustris-1 (open red squares) and the Illustris-TNG-100 (blue dots) samples at $z=0$ and comparison of the models with the WINGS data (black dots). There are 2400 objects for the Illustris-1 sample and about 600 objects for the Illustris-TNG-100. The mean radii of Illustris-1 are smaller than about a factor of two for stellar masses smaller that about $6\, 10^{10}\, M_\odot$, while they are nearly equal if not slightly larger above this limit. Right panel: The \Lsig\ plane of the same data. The symbols and color codes are the same as in the left panel.  }
              \label{fig_1}
    \end{figure*}

A detailed analysis of the differences between Illustris-1 and Illustris-TNG data has been made by \citet{Pillepich_etal_2018a,Pillepich_etal_2018b}, \citet{Rodriguez-Gomez_etal_2019}, and \citet{Huertas-Company_etal_2019}. One of the issues of major tension between the two suites of models concerns the radii of the low mass galaxies (roughly of $M_s \leq 5\,10^{10}\,M_\odot$ where the Illustris-TNG radii are about a factor of two smaller that those of Illustris-1  while above it they are nearly equal \citep{Pillepich_etal_2018a,Pillepich_etal_2018b,Rodriguez-Gomez_etal_2019}. 
\cite{Huertas-Company_etal_2019} compared the \MR\ plane built with the two sources above and the SDSS data of \citet{Meert_etal_2015} finding the same result (see their Fig. 11).

To better illustrate the difference in Fig. \ref{fig_1} we compare the data of Illustris-1 with those of Illustris-TNG-100 and the WINGS objects. The difference in the low mass range is confirmed, but the hockey-stick like shape of the distribution of model galaxies in the two samples is the same (see also Fig. \ref{fig_9} and \ref{fig_11} below). 

In addition to this, there is the claim that Illustris-1 simulations do not produce a realistic red sequence of galaxies due to insufficient quenching of the star formation with too few red galaxies   \citep{Snyder_etal_2015, Bottrell_etal_2017a, Bottrell_etal_2017b,Nelsonetal2018,Rodriguez-Gomez_etal_2019}, while the Illustris-TNG simulations produce a much better red sequence \citep{Nelsonetal2018, Rodriguez-Gomez_etal_2019}. There is also the problem of the insufficient number of red galaxies with respect to the observed population of ETGs. This is of little importance for our analysis because we do not make use of colors but only of total luminosities. 

Concerning the internal structure of the Illustris-1 galaxies, \citet{Bottrell_etal_2017b} measured the Sersic index, the axis ratio and the radii of these galaxies and found that too few bulge-dominated objects are produced in tension with observations. In contrast the Illustris-TNG galaxies have much better internal structural parameters \citep{Rodriguez-Gomez_etal_2019}. Fortunately, the point mass view of the Illustris-1 models we have adopted secures that our analysis is not too much affected by this problem.

Finally, the Illustris-1 data-set does not give information about the morphology of the galaxies. This means that in our comparison ETGs and late-type objects are mixed in our plots. Again Fig. 11 of \cite{Huertas-Company_etal_2019} shows us that ETGs and late-type objects follow very similar trends in the Sc-Rs. The basic features of the Sc-Rs shown by ETGs are not destroyed with the addition of late-type objects. 

Anyway, since in our work we do not make any prediction, but only qualitatively compare observations and simulations. Looking at their behavior in the FP projections, the good match between data and simulations, and the fact  model galaxies  are able to reproduce some particular features visible in  the FP projections (like e.g. the position of the BCGs and the existence of a ZoE) lend support to the scenario proposed here. All this  makes us confident that the simulations produce galaxies with luminosity and primary structural parameters not too far from those of real galaxies.

Given the large heterogeneity of data used here, we remark that
the completeness of the data sample is not fundamental for the conclusions drawn in this work, because we neither make any statistical analysis of the data nor we fit any distribution. The data are only used qualitatively to show that our calculations are in agreement with the observed distributions of ETGs in the main Sc-Rs. 

The purpose of this paper is only that of proposing a new possible framework to analyze  the evolution of ETGs.

\section{The equations of galaxy evolution}
\label{sec:3}

{The equations tracking the evolution of galaxies are based on two hypotheses: 1) ETGs are always close to the virial equilibrium, a reasonable assumption since the dynamical time scale to reach such condition is of the order of the free-fall time ($<300$ Myrs); 2) the \Lsigb\ law somehow mirrors  the effects of many internal and external events affecting luminosity and velocity dispersion (\ie, mass).
The two equations are:

\begin{eqnarray}
 \sigma^2 &= & \frac{G}{k_v} \frac{M_s}{R_e}  \\
 \sigma^\beta &= & \frac{L}{L'_0} = \frac{2\pi I_e R^2_e}{L'_0} 
\label{eqsig}
\end{eqnarray}
 where $k_v$ is the non homology parameter defined by \cite{Bertinetal2002}. The unknown variables of this system of equations to be found are $\beta$ and $L'_0$. 

Combining these two equations are together, one can write:}

\begin{equation}
    a_1 \log\sigma + b_1\log I_e + c_1 \log R_e + d_1 = 0
    \label{eqfege}
\end{equation}

\noindent
where the coefficients:

\begin{eqnarray}
a_1 & = & \beta-2 \\ \nonumber
b_1 & = & -1 \\ \nonumber
c_1 & = & -3 \\ \nonumber 
d_1 & = & \log(M_s)-\log(k_v/G)-\log(2\pi/L'_0) \nonumber
\end{eqnarray}

\noindent
are written in terms of $\beta$ and $L'_0$.
The similarity with the FP equation is clear. This is the equation of a plane in the \FPR\ space. The novelty is that each galaxy follows independently an equation like this. In this case, since $\beta$ and $L'_0$ are time dependent, the equation is telling us which is the instantaneous direction of motion of an object in the \FPR\ space and in its projections.

Before showing this, let us trace back the past history of the reasoning presented in this section.  Starting from the same arguments and equations (\ref{eqsig} and \ref{eqfege}), after tedious algebraic manipulations \citet{Donofrio_Chiosi_2022} arrived to a cubic equation in the variable $\beta$ (their eqn. 10), the coefficients of which where function of $\sigma$, $I_e$, $R_e$, $M_s$, and $L$. The cubic equation was applied to real galaxies of the  WINGS list and model galaxies of the Illustris-1 catalog. In most cases three real roots were found, two of them positive and one negative.  In some cases the solutions were complex and this was attributed to insufficient accuracy in the input parameters. The mutual agreement between the two sets of data (WINGS and Illustris-1) was considered as a strong hint for self consistency of the whole approach.  This agreement was indeed misleading because it masked first an algebraic mistake made while carrying out  the lengthy analytical manipulations (i.e. a factor 0.5 missing in front of a group of terms in logarithmic form), second that the agreement between WINGS and Illustris-1 made via the cubic equation was in reality a circular {argument as in each case the results would have been the same regardless of whether the equation was correct or not.} Furthermore,  attempts to incorporate the cubic equation in model galaxies did not lead to a clear understanding of the physical role and meaning played by the three \Lsigb\ relations associated to each time step (the factor $L'_0$ being derived from the real luminosity by comparison). It was clear that some of the $\beta$s changed sign in the course of evolution and also that complex solutions could occur during the lifetime of a galaxy, the low mass ones in particular. However, from these results the tantalizing suggestion came out  that a solution of the puzzle could be reached by changing strategy. All this led us to revise the whole problem thus discovering the analytical mistake and putting  the mathematical formulation on the right track. The new version of the problem is presented here below. The cubic is replaced by a system of equations in the unknowns $\beta$ and $\log L'_0$, thus fully determining the \Lsigb\ and its evolutionary history. 

Starting from eqs. \ref{eqsig} and \ref{eqfege}, after some algebra it is possible to write all the relations among the parameters of the FP projections. For the \IeRe\ plane we have:

\begin{equation}
I_e  = \Pi R_e^{\gamma}
\label{eqIeRe}
\end{equation}
\noindent
where 

$$\gamma=\frac{(2/\beta)-(1/2)}{(1/2)-(1/\beta)}$$

\noindent
and
$\Pi$ is a factor that depends on $k_v$, $M/L$, $\beta$, and $L'_0$ and id described by:

$$
\Pi  = \left [ \left (\frac{2\pi}{L'_0}\right )^{1/\beta} \left (\frac{L}{M_s} \right )^{(1/2)} \left (\frac{k_v}{2\pi G} \right )^{(1/2)} \right ]^{\frac{1}{1/2-1/\beta}}.
\label{eq4}
$$

\noindent
For the \Rsigma\ plane we have:

\begin{equation}
    R_e = \left [ \left (\frac{k_v}{G}\right ) \left (\frac{L'_0}{2\pi}\right ) \left (\frac{1}{M_s} \right ) \left (\frac{1}{I_e}\right ) \right ] \sigma^{(2+\beta)},
    \label{eqReSig}
\end{equation}

\noindent
for the \IeSig\ plane:

\begin{equation}
    I_e = \left [ \left (\frac{G}{k_v}\right ) \left (M_s \right ) \left (\frac{L'_0}{2\pi}\right )
    \left (\frac{1}{R^3_e} \right )\right ] \sigma^{(\beta-2)}
    \label{eqIeSig}
\end{equation}

\noindent
and for the \MRa\ plane:

\begin{equation}
R_e = \left [ \left (\frac{G}{k_v}\right ) \left (\frac{L'_0}{2\pi}\right )^{2/\beta} \left (\frac{1}{I_e}\right )^{2/\beta} \right ]^{\beta/(\beta+4)} M_s^{\beta/(\beta+4)}.
\label{eqRM}
\end{equation}

\begin{figure}
   \centering
   \includegraphics[scale=0.42]{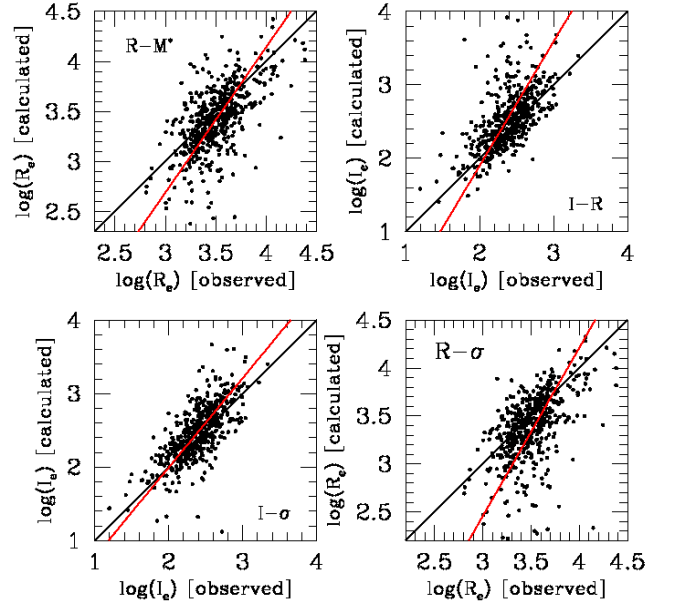}
   \caption{Comparison between observed and calculated parameters. The data of the WINGS database are used. The black solid line marks the 1:1 relationship. The red solid line is the bi-linear least square fit of the distribution.}
              \label{fig_2}
    \end{figure}

It should be remarked here that these equations do not represent the true physical link between two variables because  their proportionality factor contains other variables as well. In other words, they do not tell us how   $R_e$ and $I_e$ vary when $\sigma$ changes.  They are  intermediate mathematical expressions yielding  the structural parameters $R_e$ or $I_e$ as functions of the others.  
Figure \ref{fig_2} gives an idea of the degree of precision in reproducing the structural parameters when eqs. (\ref{eqIeRe}), (\ref{eqReSig}), (\ref{eqIeSig}) and (\ref{eqRM}) are used. The x-axis contains the measured parameters, while the y-axis the values calculated on the basis of our equations. The scatter in log units ranges from $0.3-0.4$, so a factor of $2-2.5$ uncertainty is possible and likely attributable to the $\sim20\%$ errors of the scaling parameters.

The importance of these equations is that, starting from them, one can also write the following equations (in log form):

\begin{eqnarray}
\beta [\log(I_e)+\log(G/k_v)+\log(M_s/L)+\log(2\pi)+\log(R_e)] + \\ \nonumber
    + 2\log(L'_0) - 2\log(2\pi) - 4\log(R_e) = 0 \\ 
 \beta\log(\sigma) + \log(L'_0) + 2\log(\sigma) + \log(k_v/G) - \log(M_s) + \\ \nonumber
 - \log(2\pi) - \log(I_e) - \log(R_e) = 0 
\label{eqbet}
\end{eqnarray}

and assuming:

\begin{eqnarray}
A  & = & \log(I_e)+\log(G/k_v)+\log(M_s/L)+\log(2\pi)+ \\ \nonumber
   &   & \log(R_e)  \\ \nonumber
B  & = & - 2\log(2\pi) - 4\log(R_e)  \\ \nonumber
A' & = &  \log(\sigma)  \\ \nonumber
B' & = & 2\log(\sigma) - \log(G/k_v) - \log(M_s) - \log(2\pi) -  \\ \nonumber 
   &   & \log(I_e) - \log(R_e)  \\ \nonumber
\end{eqnarray}
write the following system:

\begin{eqnarray}
A\beta + 2\log(L'_0) + B = 0 \\ \nonumber
A'\beta + \log(L'_0) + B'= 0
\label{eqsyst}
\end{eqnarray}

with solutions:

\begin{eqnarray}
 \beta & = & \frac{-2\log(L'_0) - B}{A} \\ \nonumber 
 \log(L'_0) & = &\frac{A'B/A - B'}{1-2A'/A}.
 \label{eqbeta}
\end{eqnarray}

In other words, it is possible to derive the values of $\beta$ and $L'_0$ for each galaxy. This means that the knowledge of the structural parameters reveals the basic step of galaxy evolution encoded in the parameters $\beta$ and $L'_0$.

   \begin{figure}
   \centering
   \includegraphics[scale=0.42]{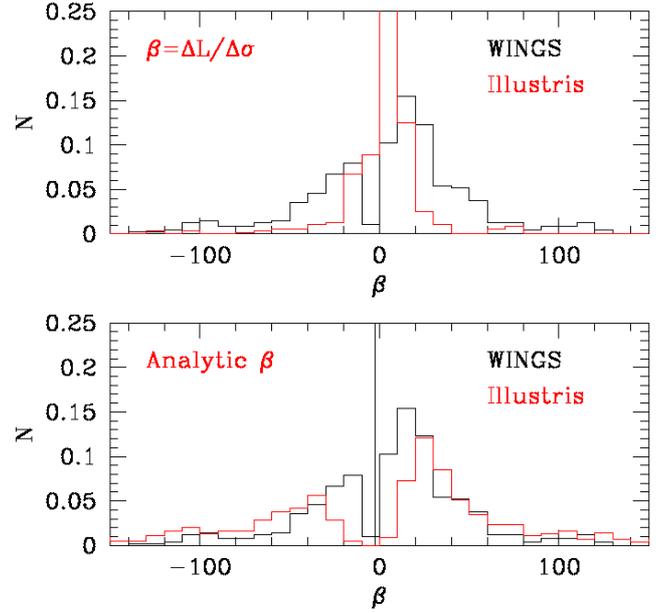}
   \caption{Upper panel: Histogram of the $\beta$ solutions derived from eq. \ref{eqbeta} for the real ETGs (black line) compared with the distribution derived for the galaxies of the Illustris-1 simulation when $\beta$ is calculated looking at the variation of luminosity and velocity dispersion in two close redshift epochs ($z=0.2$ and $z=0$); Lower panel: Histogram of the $\beta$ solutions derived from eq. \ref{eqbeta} for the real ETGs (black line) and the Illustris-1 galaxies (red line)
   that are close to the virial equilibrium. The solid black line marks the average value of $\beta$.}
              \label{fig_3}
    \end{figure}
    
Fig. \ref{fig_3} shows the histograms of the distributions of the $\beta$ parameter derived for the galaxies of the WINGS and Illustris-1 samples respectively. {In the upper panel the $\beta$ values for the Illustris-1 data are obtained from the $\Delta L/\Delta\sigma$ ratio measured on the \Lsig\ plane. This is possible by considering the values of $L$ and $\sigma$ at two close redshift epochs ($z=0.2$ and $z=0)$. In the lower panel we have  considered only the objects that are close to the virial equilibrium, i.e. those for which:

\begin{equation}
2\log(\sigma)=\log(G/k_v)+\log(M_s/L)+\log(2\pi)+\log(I_e)+\log(R_e) 
\label{eqDegV}
\end{equation}
within a 20\% uncertainty, and calculated $\beta$ using our new analytical equations.} When this condition is satisfied we get that $2A'/A=1$ and $\beta$ and $L'_0$ diverge.

Notably the values of $\beta$ are both positive and negative and there is a clear deficiency of objects with $\beta$ close to 0. This is true both for WINGS and Illustris-1. 
{The average value of $\beta$ is $-2.44$ with a rms scatter of $\sim178$. The positive values range from 1.05 to 1531, while the negative ones from $-5.4$ to $-3860$.}

The importance of {eqs. (\ref{eqbeta}) } is that we have now an empirical thermometer of the virial condition, realized when $\beta$ and $L'_0$ diverge.

   \begin{figure}
   \centering
   \includegraphics[scale=0.40]{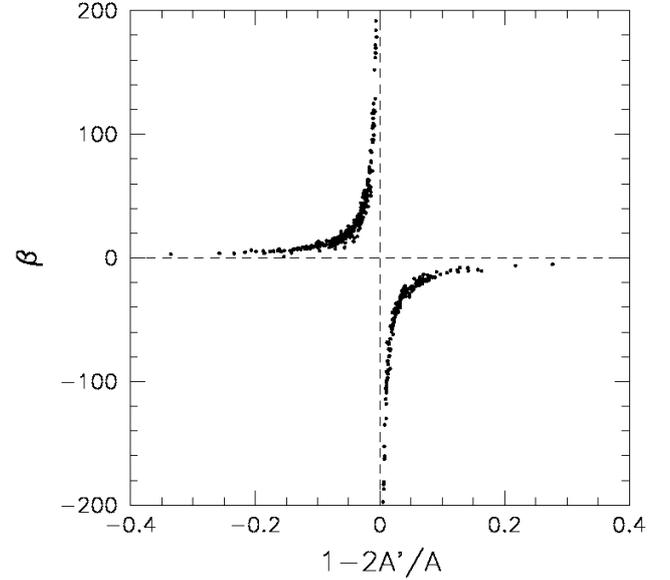}
   \caption{The distribution of $\beta$ as a function of the degree of virialization.}
              \label{fig_4}
    \end{figure}

The meaning of Fig. \ref{fig_3} is that galaxies during their evolution can acquire either positive and negative values of $\beta$, depending on the particular events experienced (merging, stripping, star formation, etc.), and this has immediate effects on the structural parameters in the Sc-Rs, that change accordingly. Consequently, the Sc-Rs seen in their temporal framework become sources of information for the global evolution of the stellar systems.   

Figure \ref{fig_4} shows the distribution of $\beta$ as a function of the degree of virialization, expressed by the quantity $1-2A'/A$ derived. The large values of $\beta$ are attained by objects very close to the virial condition.
On the other hand, the small $\beta$'s belong to objects still away from this condition.

   \begin{figure}
   \centering
   \includegraphics[scale=0.42]{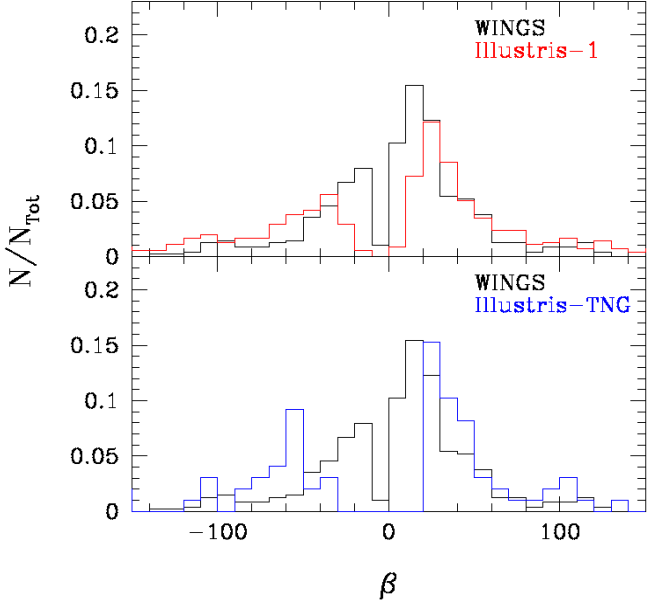}
   \caption{Histogram of the $\beta$ solutions derived from eq. \ref{eqbeta} for the real ETGs (black line) of WINGS,  the Illustris-1 galaxies (red line) that are close to the virial equilibrium, and the Illustris-TNG galaxies (blue line) in the same conditions.  The three histograms are qualitatively similar, thus confirming that our analysis depends little on the choice between the two theoretical data-bases.}
              \label{fig_5}
    \end{figure}

In closing this section, we show the distribution of the $\beta$s we would obtain with the galaxy models of the Illustris-TNG-100 sample and compare it with those of Illustris-1 and the WINGS data. The three histograms  are shown in Fig. \ref{fig_5}. The difference is very small and largely due to the smaller number of galaxies in the Illustris-TNG sample. 

\subsection{Trends in the FP projections }

In this section we try to better explain the reasons why $\beta$ can change sign during the life of a galaxy or when passing from one galaxy to another.  The advantage of knowing  $\beta$ is much clear when we look at the projections of the FP. Eqs. (\ref{eqIeRe}), (\ref{eqReSig}), and (\ref{eqIeSig}) can be further elaborated to eliminate the dependence on $I_e$ and $R_e$ present in their zero-points. We get: 

\begin{equation}
    I_e = \left [ \frac{G}{k_v}\frac{L'_0}{2\pi}M_s \Pi^{3/\gamma} \right ]^{\frac{\beta-2}{1+3/\gamma}} \sigma^{\frac{\beta-2}{1+3/\gamma}}
    \label{relIeSig}
\end{equation}

\begin{equation}
    R_e = \left [ \frac{G}{k_v}\frac{L'_0}{2\pi}\frac{M_s}{\Pi} \right ] \sigma^{\frac{\beta-2}{3+\gamma}}
    \label{relReSig}
\end{equation}

\begin{equation}
    R_e = \left [ (\frac{G}{k_v})^{\beta/2} \frac{L'_0}{2\pi} \frac{1}{\Pi} \right ]^{\frac{2(\beta-2)}{\beta^2-6\beta+12}} M_{s}^{\frac{\beta^2-2\beta}{\beta^2-6\beta+12}}
    \label{relReM}
\end{equation}

\begin{figure}
   \centering
   \includegraphics[scale=0.42]{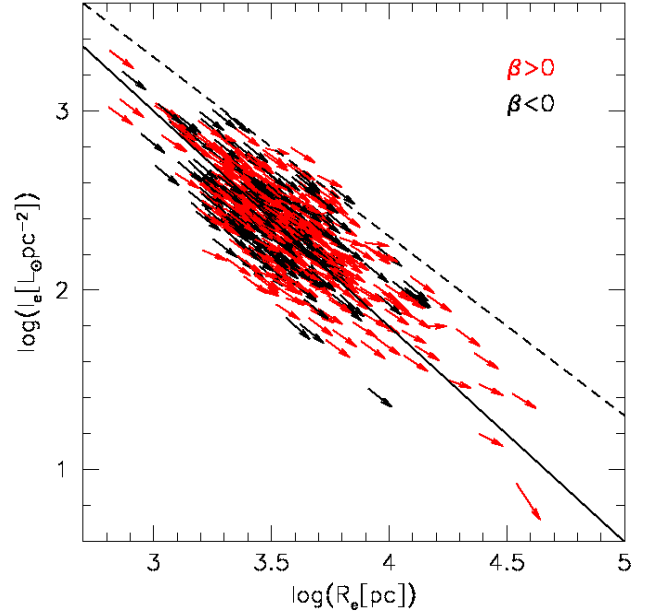}
   \caption{The \IeRe\ plane. The black and red arrows mark the direction of motion of galaxies in this plane for large negative and positive values of $\beta$. The black solid line gives the lsq fit of the data, while the broken line represents the zone of exclusion.}
              \label{fig_6}
    \end{figure}

\begin{figure}
   \centering
   \includegraphics[scale=0.42]{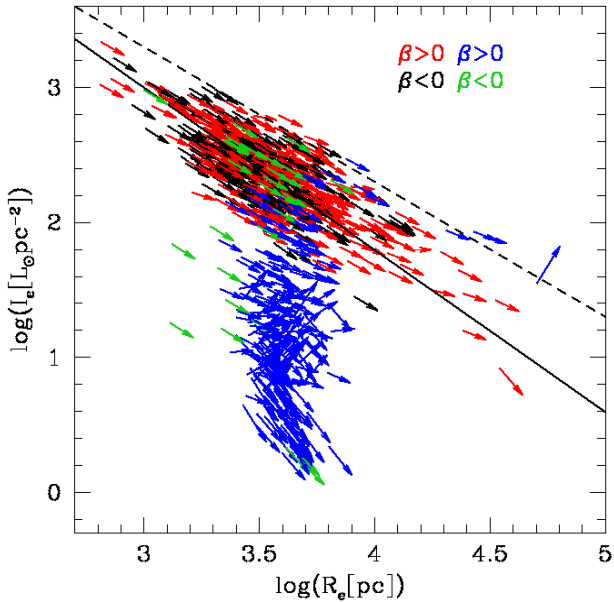}
   \caption{The \IeRe\ plane for the WINGS and Illustris-1 data. The black and red arrows mark the direction of motion of the WINGS galaxies in this plane for large negative and positive values of $\beta$. The green and blue arrows are those of Illustris-1 for negative and positive values of $\beta$. {In the plot we have used only 1/10 of the Illustris-1 galaxies in order to permit to distinguish the objects with different $\beta$ values moving in different directions.}}
              \label{fig_7}
    \end{figure}

\begin{figure}
   \centering
   \includegraphics[scale=0.42]{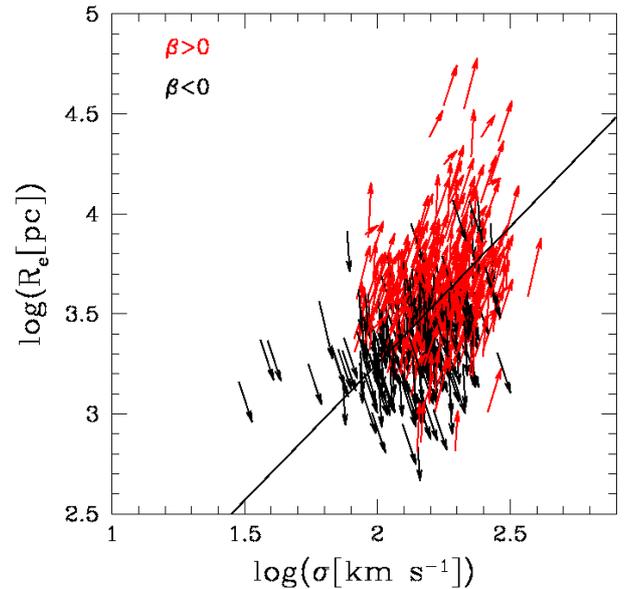}
   \caption{The \Rsigma\ plane. Symbols and colors as in Fig. \ref{fig_6}.}
              \label{fig_8}
    \end{figure}

\begin{figure}
   \centering
   \includegraphics[scale=0.42]{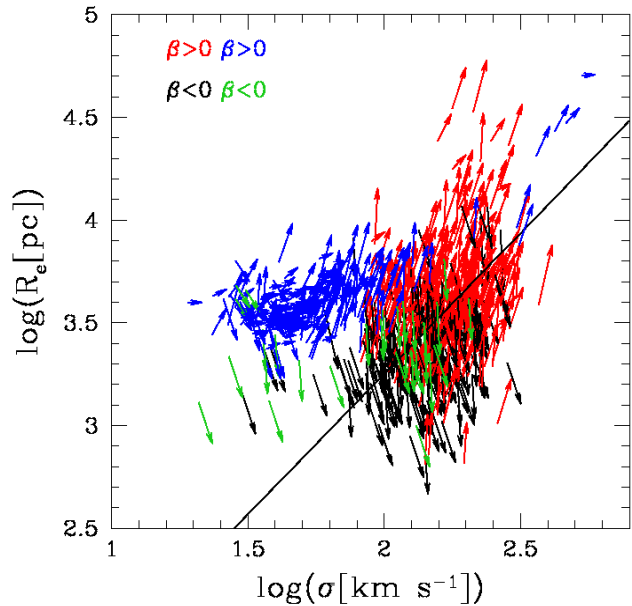}
   \caption{The \Rsigma\ plane for WINGS and Illustris-1. Symbols and colors as in Fig. \ref{fig_7}.}
              \label{fig_9}
    \end{figure}

\begin{figure}
   \centering
   \includegraphics[scale=0.42]{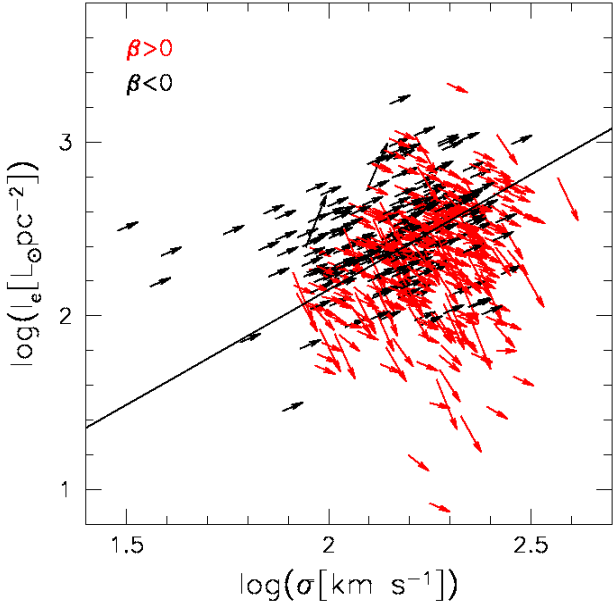}
   \caption{The \IeSig\ plane. Symbols and colors as in Fig. \ref{fig_6}.}
              \label{fig_10}
    \end{figure}

\begin{figure}
   \centering
   \includegraphics[scale=0.42]{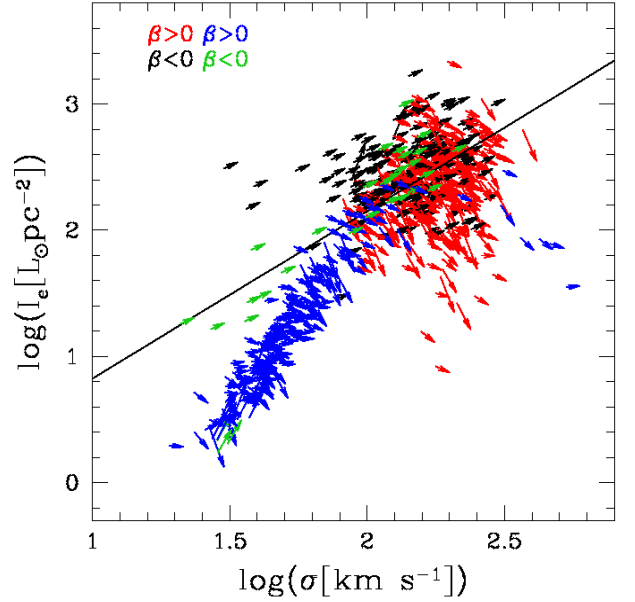}
   \caption{The  \IeSig\ plane for WINGS and Illustris-1. Symbols and colors as in Fig. \ref{fig_7}.}
              \label{fig_11}
    \end{figure}

These relations now better represent the mutual dependence of the structural parameters (e.g. of $I_e$ as a function of $R_e$ and $\sigma$ and of $R_e$ as function of $\sigma$) and clarify what is the role of $\beta$. When a galaxy moves in the \Lsig\ plane, according to the values of $\beta$, it does the same also  in the other FP projections, according to the slopes reported in the last three columns of Tab. \ref{Tab_1}. These slopes depend on $\beta$ and indicate the direction of motions (marked by the arrows) that are visible in Figs. \ref{fig_6}-\ref{fig_11}.

Figure \ref{fig_6} shows the case of the \IeRe\ plane. The black and red arrows mark the direction of motion of galaxies predicted on the basis of their negative and positive values of $\beta$ respectively. Since the WINGS galaxies are well virialized, the values of $\beta$ are always very large, either positive and negative. Both such slopes give consistently a direction of motion close to $\sim-1$ in the \IeRe\ plane. The $-1$ slope is that predicted on the basis of the VT (represented by the broken line which also marks the ZoE) \citep[see,][]{DonofrioChiosi2021}). We note that no galaxies can cross the ZoE, because their motion is nearly parallel to the ZoE.

These arguments  demonstrate the reason  why there is a ZoE in the \IeRe\ plane: the only possible direction of motion for well virialized objects is that with  slope $\sim-1$ generated by the large positive and negative values of $\beta$. 

When $\beta\sim 0$, \ie\ when the galaxies are less virialized, they can move in other directions in this plane. 
Unfortunately, our sample does not include the dwarf ETGs, that are usually distributed in a cloud, below the ZoE with radii lower than 3-4 kpc \citep[see e.g.][]{Capacciolietal1992, Donofrioetal2020}. For these objects we predict values of the slopes in all possible directions.  

The only way to check this is to make use of the model galaxies  either of Illustris-1 or Illustris-TNG. Figure \ref{fig_7} confirms our prediction. Although the data of Illustris-1 are affected by the well known problem of the systematically larger \re\ with respect to the observed ones \citep{Donofrioetal2020, Bottrelletal2017}, we can note that  several objects have arrows nearly orthogonal to those of the well virialized galaxies. The expected motions of the dwarf galaxies are in all possible directions, thus giving rise to the cloud of the "ordinary" ETGs defined by \cite{Capacciolietal1992}.  Furthermore, when the distribution curves its shape, we note a progressive variation of the arrow directions. This means that the overall distribution in this plane is governed by the different movements of the galaxies in the \Lsig\ plane described empirically by the different values of $\beta$ and $L'_0$.

The direction of the arrows displayed in each figure visualizes the expected displacement of a galaxy based on the actual value of $\beta$. However, the arrows only give the direction of motion, not the orientation of the future temporal evolution of a galaxy. Furthermore, they do not indicate the path followed by each galaxy to reach the current observed position in the diagrams.

\begin{table*}
\begin{center}
\caption{The slopes of the \IeRe, $R_e-\sigma$, \IeSig\ and  $R_e-M_s$ planes for different values of $\beta$.}
                \label{Tab_1}
                \begin{tabular}{|r| r| r| r| r| r| r| r|}
\hline
         $\beta$     & \IeRe\ & $R_e-\sigma$ $^a$ & \IeSig\ $^b$ & $R_e$-$M_s$ $^c$ &  $R_e-\sigma$ $^d$ & \IeSig\ $^e$ & $R_e$-$M_s$ $^f$  \\
\hline
       100.0 &      -0.98   &    102.0  &  98.0 &  0.96 & 48.50 & -47.51 & 1.04 \\
        50.0 &      -0.96   &     52.0  &  48.0 &  0.92 & 23.51 & -22.53 & 1.08 \\
        10.0 &      -0.75   &     12.0  &   8.0 &  0.71 &  3.55 &  -2.66 & 1.54 \\
         5.0 &      -0.33   &      7.0  &   3.0 &  0.55 &  1.12 &  -0.37 & 2.14 \\
         3.0 &       1.00   &      5.0  &   1.0 &  0.43 &  0.25 &   0.25 & 1.00 \\
         2.0 &       0.00   &      4.0  &   0.0 &  0.33 &  0.0  &    0.0 & 0.00 \\
         1.0 &      -3.00   &      3.0  &  -1.0 &  0.20 &  0.0  &    0.0 &-0.14 \\
         0.5 &      -2.33   &      2.5  &  -1.5 &  0.11 & -2.25 &   5.25 &-0.08 \\
         0.0 &      -2.00   &      2.0  &  -2.0 &  0.00 & -2.00 &   4.00 & 0.28 \\
        -0.5 &      -1.80   &      1.5  &  -2.5 & -0.14 & -2.08 &   3.74 & 0.08 \\
        -1.0 &      -1.67   &      1.0  &  -3.0 & -0.33 & -2.25 &   3.75 & 0.16 \\
        -2.0 &      -1.50   &      0.0  &  -4.0 & -1.00 & -2.66 &   4.00 & 0.28 \\
        -3.0 &      -1.40   &     -1.0  &  -5.0 & -3.00 & -3.12 &   4.37 & 0.38 \\
        -5.0 &      -1.28   &     -3.0  &  -7.0 &  5.00 & -4.08 &   5.25 & 0.52 \\
       -10.0 &      -1.16   &     -8.0  & -12.0 &  1.67 & -6.54 &   7.63 & 0.69 \\
       -50.0 &      -1.03   &    -48.0  & -52.0 &  1.08 &-26.51 &  27.53 & 0.92 \\
      -100.0 &      -1.02   &    -98.0  &-102.0 &  1.04 &-51.50 &  52.51 & 0.96 \\
\hline
\end{tabular}
\end{center}
Notes: a) Slope when $k_v$, $M_s$ and $I_e$ are constant; b) Slope when $k_v$, $M_s$ and $R_e$ are constant; c) Slope when $k_v$, and $I_e$ are constant; d) Slope when $k_v$ and $M_s$ are constant: e) Slope when $k_v$ and $M_s$ are constant; f) Slope when $k_v$ is constant.
\end{table*}

The same can be said for the other two FP projections.
Figures \ref{fig_8} and \ref{fig_10} represent the \Rsigma\ and \IeSig\ planes respectively. Here, the role of $\beta$  is much more clear. It is well evident in fact that the galaxies with negative $\beta$'s move in different directions with respect to those with positive $\beta$, originating the curvatures observed in these diagrams.

Once more the addition of the Illustris-1 data (Fig. \ref{fig_7} and \ref{fig_9}) confirms that the slopes derived from the $\beta$s  are consistent with the observed distribution of ETGs and demonstrates that the observed curvatures originate from the different motion of galaxies with positive and negative values of $\beta$.

\begin{figure}
   \centering
   \includegraphics[scale=0.42]{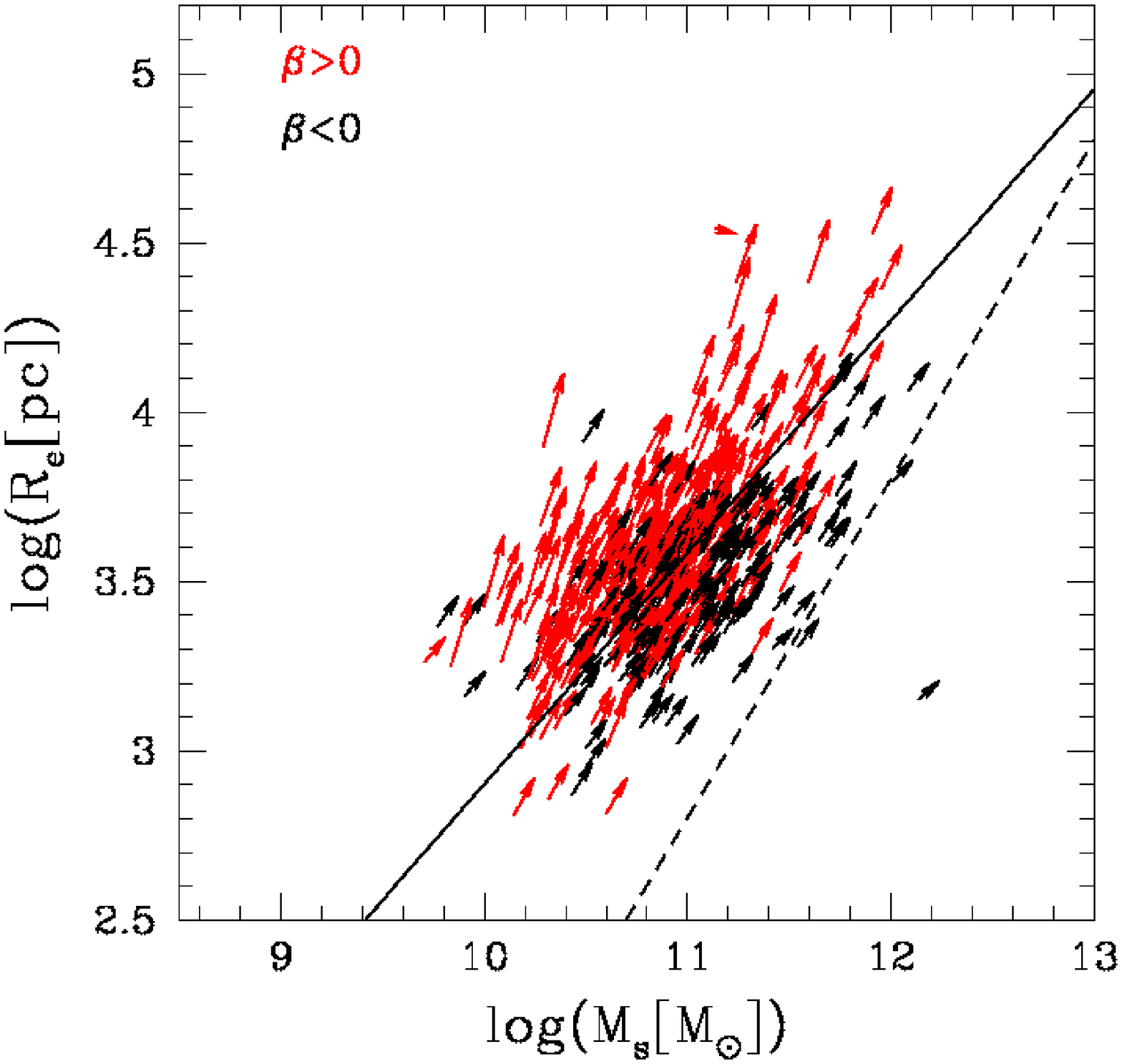}
   \caption{The \MR\ plane for the WINGS galaxies. Symbols and colors as in Fig. \ref{fig_6}.}
              \label{fig_12}
    \end{figure}

\begin{figure}
   \centering
   \includegraphics[scale=0.42]{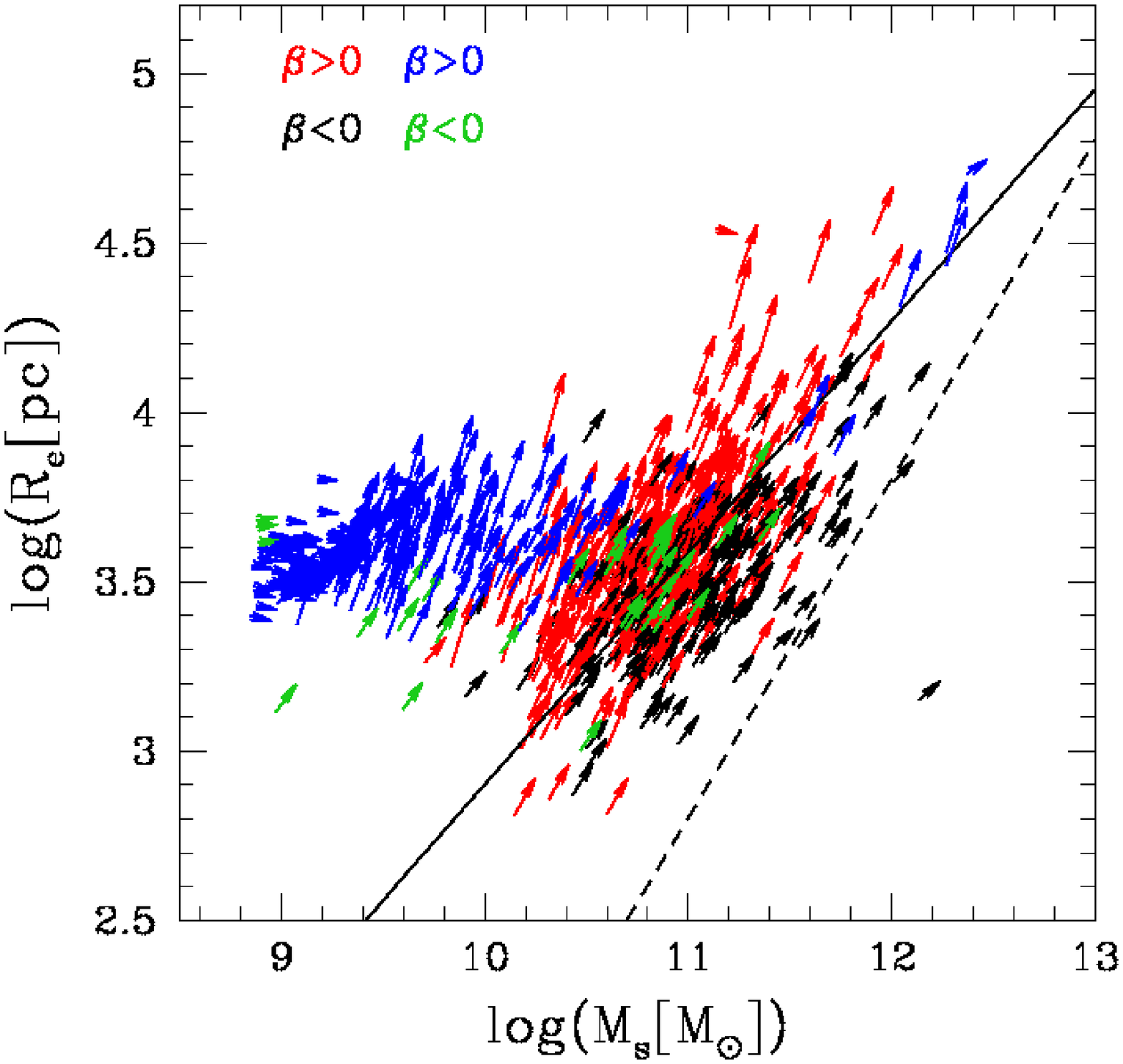}
   \caption{The \MR\ plane for WINGS and Illustris-1 galaxies. Symbols and colors as in Fig. \ref{fig_7}.}
              \label{fig_13}
    \end{figure}

Figures \ref{fig_12} and \ref{fig_13} are similar plots for the \MR\ plane. Even in this important diagram we observe a ZoE (marked by the dashed line with slope equal to 1). Our {calculations} predict why this ZoE is here: the reason is that all the virialized objects (with large $\beta$ values) can only move in the direction with slope equal to 1 (see Table \ref{Tab_1}). The \re\ of Illustris-1 are notoriously somewhat larger than those measured, but the general behavior is in good agreement with the observed distribution. The galaxies with large positive and negative values of $\beta$ move with a slope close to 1, while in the cloud of points with small masses we can note objects with different directions.

We conclude that all the positions of ETGs in the FP projections and in the \MR\ plane depend on the motions occurred during the peculiar evolutionary path followed by {each galaxy}. When the galaxies are well virialized these motions can occur only in well fixed directions depending on the value of $\beta$.

This is a coherent and self-consistent explanation of all the main scaling relations   built with  the structural parameters of ETGs. It follows that even the FP, the father of the scaling relations,  must find a similar explanation.

Looking at Table \ref{Tab_1} in detail we also note that:
\begin{itemize}
    \item In all FP projections, when $\beta$ becomes progressively negative, \ie\ when the objects are rapidly declining in their luminosity at nearly constant $\sigma$, the slopes either converge to the values predicted by the VT (in the \IeRe\ relation and in the \MRa\ relation), or diverge toward large values (in the $I_e-\sigma$ and $R_e-\sigma$ relations), because the galaxy keeps its velocity dispersion when the luminosity decreases (only \Ie\ and \re vary). This offers a natural explanation of the ZoE.
    \item Positive and negative values of $\beta$ are equally permitted, both with real and simulated data. In general, the objects that are still active in their star formation or have recently experienced a merger, have positive values of $\beta$, while those progressively quenching their SF have increasing negative $\beta$.   
    \item The "curvature" in the observed distributions (\ie\ the transition from the large cloud of  small galaxies to the much narrow tail of the brightest objects) is naturally explained by the existence of positive and negative values of $\beta$.
\end{itemize}

\begin{table}
\begin{center}
\caption{Trends of the effective parameters as a consequence of changes in $\sigma$ and $L$.}
                \label{Tab_2}
                \begin{tabular}{|l| |r| |r| |l|}
\hline
        \multicolumn{4}{|c|}{}   \\
        \multicolumn{4}{|c|}{$\beta>0$}   \\
        \multicolumn{4}{|c|}{}   \\
\hline
$L \& \sigma \nearrow$ & $R_e \nearrow$ & $I_e (const.\, or \searrow)$ &$ M_s \nearrow$ \\
$L \& \sigma \searrow$ & $R_e \searrow$ & $I_e (const.\, or \nearrow)$ &$ M_s \searrow$ \\
\hline
        \multicolumn{4}{|c|}{}   \\
        \multicolumn{4}{|c|}{$\beta<0$}   \\
        \multicolumn{4}{|c|}{}   \\
\hline
$L \searrow \& \, \sigma \nearrow$ & $R_e \searrow$ & $I_e (const.\, or \nearrow)$ &$ M_s (const. \, or \nearrow)$ \\
$L \nearrow \& \, \sigma \searrow$ & $R_e \nearrow$ & $I_e (const.\, or \searrow)$ &$M_s (const. \, or \searrow)$ \\
\hline
\end{tabular}
\end{center}
\end{table}

A way to better understand the effects played by $\beta$ is to think at the possible variations of \re\ and \Ie\ when $L$ and $\sigma$ vary in the \Lsig\ plane. There are four possible changes of $L$ and $\sigma$ in this plane. They are schematically shown in Table \ref{Tab_2}, which displays, according to the values of $\beta$,  the expected variations of \re and \Ie, when $L$, $M_s$ and $\sigma$ vary. Note that when $\beta$ is negative, not necessarily there is a decrease in luminosity, and when $\beta$ is positive, a decrease in luminosity might also occurs.

When the luminosity of a galaxy changes, both the effective radius and the mean effective surface intensity \Ie\ vary. This happens because $R_e$ is not a physical radius, like e.g. the virial radius (which depends only on the total mass), but it is the radius of the circle that encloses half of the galaxy total luminosity. Since the ETGs have different stellar populations with different ages and metallicity, it is highly improbable that the decrease in luminosity does not change the whole appearance of the luminosity profile\footnote{This could happen only in a coeval stellar system with the same type of stars in any galaxy volume.}. Consequently the growth curve changes and determines a variation of $R_e$ and $I_e$.
If the luminosity decreases passively, in general one could expect a decrease of $R_e$ and an increase of $I_e$. On the other hand, if a shock induced by harassment or stripping induces an increase of $L$ (and a small decrease in $\sigma$), we might expect an increase of $R_e$ and a decrease of $I_e$. 

The observed variations of these parameters depend strongly on the type of event that a galaxy is experiencing (stripping, shocks, feedback, merging, etc.).
In general, one should keep in mind that these three variables $L$, $R_e$ and $I_e$ are strongly coupled each other and that even a small variation in $L$ might result in  ample changes of $R_e$ and $I_e$.
In this context, we begin to understand that the Sc-Rs are useful tools for guessing both the dynamics and the evolutionary state of the stellar content of a galaxy.

In summary, what we claim here is that all the above diagrams should be analyzed taking into account the effects of time and should not be investigated separately. They are snapshots of an evolving situation, and such temporal evolution cannot be discarded.  The \Lsigb\ law catches such evolution in the correct way by predicting the  direction of the future motion of each galaxy in the diagnostic planes (D'Onofrio and Chiosi, A\&A submitted). In principle, this way of reasoning should allow us to understand why galaxies are in the positions observed today in each diagram. As $\beta$ gives only the present direction of motion and not that of the motion in the past, the simultaneous use of simulations and high redshift observations might help to infer the possible precursors  of the present day galaxies on the basis of the physical properties and the distribution in the FP projections, indicated by the  values of $\beta$. In other words these scaling relations become a possible tool for inferring the evolutionary path of each galaxy.

\subsection{Origin of the  FP and its tilt}\label{FPorigin}

The final step is that related to the question of the origin of the FP. Equation (\ref{eqfege}) tells us that each galaxy follows its own FP-like equation, whose coefficients are functions of $\beta$. Starting from eq. (\ref{eqfege}) it is possible to derive the coefficients $a$, $b$ and $c$ of the plane hosting each single ETG. To do this we adopt the notation that is commonly used for the FP, in which \muem\ is expressed in mag arcsec$^{-2}$ and $R_e$ in kpc: 

\begin{equation}
 \log(R_e)=a\log(\sigma)+b<\mu>_e+c.
\end{equation}

\noindent
We get:
\begin{eqnarray}
\label{FPcoeff}
 a & = & a_1/(-c_1) \\  \nonumber
 b & = & b_1/(-c_1) \\  \nonumber
 c'& = & d_1/(-c_1) \\  \nonumber 
 \end{eqnarray}

\noindent
where $a_1$, $b_1$, $c_1$, and $d_1$  are from eq. (\ref{eqfege}), and $c=(10.56*b_1)/-c_1)-3+c'$.  This transformation is necessary because in our notation $I_e$ is expressed in $L_\odot pc^{-2}$ and $R_e$ is in pc. 

  \begin{figure}
   \centering
   \includegraphics[scale=0.42]{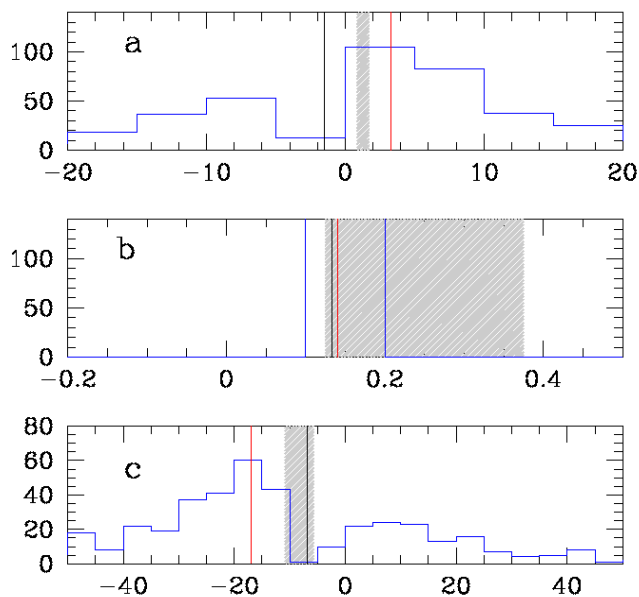}
   \caption{Histograms of  the values of the calculated FP coefficients a (top panel), b (middle panel) and c (bottom panel). In each panel we show the histogram of the coefficient (blue line), the average value (the vertical black line) and the median (the vertical red line). In the case of the $b$ coefficient, that does not depend on $\beta$, average and median are the same (the black and red vertical lines coincide). The dashed gray regions mark the intervals of the FP coefficients found by fitting the distribution of ETGs in the \FPR\ space.}
              \label{fig_14}
    \end{figure}

The distribution of these coefficients for all the WINGS sample of 479 galaxies is shown in Fig. \ref{fig_14}.
It is clear from the figure that the values for the FP coefficients, derived from the fit of the ETGs distribution and indicated by the dashed areas in each panel, are very close to the average of the single coefficients calculated by eq. \ref{FPcoeff} (the vertical black lines).
The gray bands show indeed the interval of $a$, $b$ and $c$ obtained by \citet{Donofrioetal2008} fitting the FP of ETGS separately for each cluster of the WINGS data-set.

In other words, we have framed the  FP and its tilt in a new context in which each  ETG follows its own eq. (\ref{eqfege}), namely FP, and contributes to shape the global FP (both  tilt and thickness) of the ETG population. Since the FP coefficients are obtained from a fit, it is clear that the final coefficients of the plane will be close to the average of the single values valid for each object. Some differences are expected because the final values will depend on the sample adopted (each having its own average) and from the technique used to perform the fit.

With this statement we do not mean to say that the various mechanisms invoked in the literature to explain the tilt and thickness  are incorrect. Rather, we claim that all of them can actually contribute to the average properties of the galaxy sample, giving rise to a different $\beta$ for each object.

Since its discovery, the FP has been the subject of several studies aimed at understanding why the plane is tilted with respect to the prediction of the VT ($M_s \propto R_e \sigma^2$), and why its intrinsic scatter is so small \citep[see \eg,][among many others]{Faberetal1987,Ciotti1991,Jorgensenetal1996,Cappellarietal2006,Donofrioetal2006,Boltonetal2007}. While the VT predicts $a=2$ and $b=-1$, the values coming from the fit of several samples of ETGs are systematically lower ($a \sim 1.2$) and higher ($b \sim -0.8$), and vary according to the sample used and the fitting strategy. 

Among the physical mechanisms invoked to explain the FP tilt we can find: 1) a progressive change of the stellar mass-to-light ratio ($M_s/L$) \citep[see \eg,][]{Faberetal1987,vanDokkumFranx1996,Cappellarietal2006,vanDokkumvanderMarel2007, Holdenetal2010,deGraafetal2021}; 2) structural and dynamical non-homology \citep[see \eg,][]{PrugnielSimien1997,Busarelloetal1998,Trujilloetal2004,Donofrioetal2008}; 3) dark matter (DM) content and distribution \citep[see \eg,][]{Ciottietal1996,Borrielloetal2003,Tortoraetal2009, Taranuetal2015,deGraafetal2021};  4) star formation history (SFH) and initial mass function (IMF) \citep[see \eg,][]{RenziniCiotti1993,Chiosietal1998,Chiosi_Carraro_2002,Allansonetal2009}; 5) the effects of environment \citep[see \eg,][]{Luceyetal1991,deCarvalhoDjorgovski1992,Bernardietal2003, Donofrioetal2008, LaBarberaetal2010, Ibarra-MedelLopez-Cruz2011, Samiretal2016}. 

Recent observational work has shown that variations in the $M_s/L$ ratio can account only for half of the tilt \citep[see][]{D'Eugenio_etal_2021}, with remainder being due to structural variation and possibly variations in the galaxy-averaged initial mass function of the stellar populations. Uncertainties in $M_s$ can affect the tilt if the error is mass-dependent, although this systematic uncertainty is not large enough \citep[see][]{Leja_etal_2019tre, Lower_etal_2020}. \citet{Schechter_etal_2014} using strong lensing measurement provides an independent estimate of $M_s$ but still finds  a tilt of the FP. Finally, the tilt is found in cosmological simulations \citep{Rosito_etal_2019a, Rosito_etal_2019b, deGraaff_etal_2022}. 

All these effects are in practice "involved" in our view of the problem. Indeed, since each sample of galaxies  has its own average  value of  $\beta$, because of the different history of mass accretion and luminosity evolution, it is easy to verify that systematic changes in the tilt could arise for the above mentioned reasons. When the  sample changes its average properties, a small variation of the tilt of the FP follows. This explains for instance why \cite{robertsonetal06} found that star-forming and quiescent galaxies follow different Sc-Rs, i.e. a different FP tilt, due to differences in the merger histories. 

With the  emerging of the hierarchical scenario of galaxy formation and evolution, some additional mechanisms for the FP tilt have been proposed: 1) the effects of dissipation-less merging \citep{Nipotietal2003}; 2) the gas dissipation \citep{robertsonetal06}; 3) the non regular sequence of mergers with progressively decreasing mass ratios \citep{Novak2008}; 4) the multiple dry mergers of spiral galaxies \citep{Taranuetal2015}.

Since the galaxy properties change with time, the slope of the FP is expected to change with redshift. This is confirmed by the numerical models of single galaxies, large scale cosmological simulations, and observational surveys at different redshifts, among others see \citet{Beifiori_etal_2017, Rosito_etal_2019a, Rosito_etal_2019b, Lu_etal_2019, Ferrero_etal_2021, deGraaff_etal_2022} and references. 

The most remarkable  physical feature of the FP is the  observed very small scatter, which amounts to $\approx0.05$ dex in the V-band. It seems to require a sort of fine tuning among different physical processes. The scatter has been attributed to: 1) the variation in the formation epoch; 2) the DM content; 3) the existence of metallicity or age trends; 4) the variations of the mass-to-light ratio $M/L$ \citep[see e.g.,][]{Faberetal1987, Gregg1992, GuzmanLuceyBower1993, Forbesetal1998, Bernardietal2003, Redaetal2005, Cappellarietal2006, Boltonetal2008, Gravesetal2009, Gravesetal2010, Augeretal2010, Magoulasetal2012}.

Our approach cannot predict the scatter around the FP, because this does not depend on the structural parameters, but on the properties of the stellar populations and the peculiar history of mass accretion/stripping. 

Although investigating the causes of the tilt and the small dispersion around the FP is beyond the aims of this paper, let as conclude this section with one consideration: it may be an amazing coincidence, but we note that going back to high redshift, the numerical simulations of Illustris-1 and Illustris-TNG show that the FP and the tail of the MRR persist until redshift $z\simeq 1.6$  and then disappear or is no longer so well defined  (D'Onofrio \& Chiosi, 2022;  Chiosi et al., 2022, in preparation). The concomitant appearance of the MRR and FP for ETGs more massive that about $10^{10}\, M_\odot$ maybe is a mere coincidence, but surely it is a question that must be investigated.

The galaxies on the tails of the FP and the MRR are massive ETGs whose stellar content is predominantly made of old stars or in the case of mergers with objects of smaller mass the percentage of younger stars does not alter significantly the luminosity and the colors of the basic stellar  populations. To quantify this statement
let us make the following example. At proceeding galaxy building via the hierarchical scenario, the  probability that a massive objects merge with another of similar mass becomes rarer and rarer as galaxies become more massive. Therefore, massive ETGs tends to evolve in isolation or merging objects of much smaller mass. In general, the merger of two galaxies with very different masses (e.g., $M_1/M_2 \simeq 1/10$) and some companion stellar activity  leaves the mass and  velocity dispersion nearly unchanged while the luminosity   first undergoes a burst of short duration and relative intensity proportional to the luminosity ratio $L_1/L_2$ (only slightly higher than the previous value).  This should  correspond to a nearly vertical shift on the FP of small amplitude. 
The $M/L$ ratio  either remain unchanged or slightly decrease  thus causing a little scatter of the FP. 

The opposite should occur in a merger between two galaxies of nearly equal mass, typical situation in the range of low mass galaxies. In this case, the mass and luminosity both change. If additional star formation occurs, there should be an additional increase of the total luminosity that depends on the amount of mass converted into new stars. Therefore,  the total luminosity should hardly  recover the pre-burst value, the  mass-weighted mean of the two component galaxies \citep[see Fig.11 in ][]{Tantalo_etal_2004b}. So most likely the luminosity  remains higher than before, and the $M/L$  ratio is expected to decrease. This should generate a tilt of the FP in the right direction. It is not easy to foresee the effect on the scatter. Better estimates require numerical simulations  of burst of star formation. In any case after a short time interval the maximum shift
in luminosity cannot overcome a factor of $\sim 2$ (0.3 dex).

\section{Application to model galaxies}\label{sec:4}

To lend support to the picture outlined above concerning the physical role and meaning of the \Lsigb\ relation and the role of the parameters $\beta$ and $L'_0$ without resorting to the numerical simulations of Illustri-1 and Illustris-TNG of which we have no control at all, we make use of very simple, almost analytical models of galaxy formation and evolution.  The ideal models of this type  suited to describe ETGs are those in which  the total mass increases by infall and the stars are formed according to a simple law of star formation rate (SFR) that have been developed long ago by \citet{Chiosi_1980} and extended by \citet{Tantalo_etal_1998}. The novelty here  is that we have incorporated the equations for $\beta$ and $L'_0$ (eqs. \ref{eqfege}) into the models once the luminosity, the radius and the velocity dispersion are calculated.  With the aid of these models and eqs. (\ref{relIeSig}) and (\ref{relReSig}) we have calculated the basic relationships $I_e-\sigma$ and $R_e-\sigma$, and finally made a cross-test of mutual consistency between the results from the galaxy models and the $\beta$ and $L'_0$ theory.
These  simple model of galaxy formation and evolution  were first proposed  by \citet{Chiosi_1980}, much later extended by \citet{Tantalo_etal_1998}, and recently used by \citet{Chiosi_etal_2017} to study the cosmic SFR and by \citet{Sciarratta_etal_2019} to investigate the galaxy color-magnitude diagram. Although they may look too simplistic compared to the numerical models of Illustris-1 and Illustris-TNG, yet they catch the main features of these latter and are suitable to our purposes.

In brief, a galaxy of total mass $M_G$ is made of baryonic (B) and dark matter (D), with mass $M_{B}$ and  $M_{D}$ respectively, and at any time satisfies the equation:

\begin{equation}
			M_{G}(t) = M_{B}(t) + M_{D}(t).
			\label{sum_bm_dm}
\end{equation}

At all times $M_{B}(t)$ and  $M_{D}(t)$ are in cosmological proportion, i.e. they satisfy the condition $M_{D}(t) = f_c M_{B}(t)$
where $f_c$ depends on the adopted $\Lambda$CDM  cosmological model of the Universe ($f_c \simeq 6.1$ in our case). 

The baryonic mass is supposed to be originally in form of gas, to flow in at a suitable rate and, when physical conditions allow it,  to transform into stars. With the same rate also dark matter is let flow in together with the baryonic matter to build up the total gravitational potential. Suitable prescriptions of their spatial distribution are needed to calculate the gravitational potential \citep[see][for more details]{Tantalo_etal_1998}. 

This kind of galaxy model is named "infall model", the essence of which resides in the gas accretion into the central region of the proto-galaxy at a suitable rate (driven by the timescale $\tau$)  and in the gas consumption by a Schmidt-like law of star formation. The gas accretion and consumption coupled together provide a time dependent SFR closely resembling the one resulting from N-body simulations \citep[e.g.][]{Chiosi_Carraro_2002,Merlin2006,Merlin2007,Merlin_etal_2012}.

At any time $t$ the baryonic mass $M_{B}$ is given by the sum: 

\begin{equation}
			M_{B}(t) = M_g(t) + M_s(t),
			\label{mass_gas_stars}
\end{equation}
		
\noindent 
where $M_g(t)$ is the gaseous mass {and $M_s(t)$ the  mass in stars}. At the beginning, both the gas and the star mass in the proto-galaxy are zero $M_g(t=0)=M_s(t=0)=0$. The rate of baryonic mass (and gas in turn) accretion is driven by the timescale $\tau$ according to:
		
\begin{equation}
			\frac{dM_{B}(t)} { dt} = {{M}}_{B,\tau}\exp(-t/\tau),
			\label{inf}
\end{equation}

\noindent
where ${M}_{B,\tau}$ is a constant with the dimensions of [Mass/Time] to be determined by imposing that at the galaxy age ${T_{G}}$ the total baryonic mass of the galaxy ${M_{B}(T_{G})}$ is reached:

\begin{equation}
			{M}_{B,\tau}  = \frac{M_{B}(T_{G})} {\tau [1 - \exp(- T_{G}/\tau)]}.
			\label{mdot}
\end{equation}
		
\noindent
Therefore, by integrating the accretion law, the time dependence of	 ${M_{B}(t)}$ is:

\begin{equation}
			{ M_{B}(t) =  { \frac{M_{B}(T_{G})}  {[1-\exp (-T_{G}/\tau)] }    }
				[1 - \exp(-t/\tau)]  }.
			\label{mas-t}
\end{equation}

Since dark matter flows in at the same rate of the baryonic matter, it obeys similar equations in which $M_B$ is replaced by $M_D$. However, since at any time $M_B$ and $M_D$,  are in cosmic  proportions, $M_D = f_c M_B$, the equations for $M_D$ are superfluous and the normalization on $M_B$ is enough.   The underlying hypothesis is that the presence of dark matter does not affect the evolution of the baryonic component, but for its effect on the gravitational potential energy. To this aim, some assumptions about the spatial distribution of $M_B$ and $M_D$ are needed. In other words, assuming spherical symmetry, the radii $R_B$ and $R_D$ must be specified. 

The timescale $\tau$  is related to the collapse time and the average cooling rate of the gas. Therefore,  it is expected to depend on the mass of the system. At the same time, the gas mass increases by infall and decreases by star formation.

The rate of star formation is modeled throughout the whole life of the galaxy with the \citet{Schmidt1959} law:

\begin{equation}
			\Psi(t)  \equiv \frac{dM_s}{dt} = \nu M_g(t)^k,
			\label{schmidtlaw}
\end{equation}
		
\noindent 
where $k$ regulates the dependency of the SFR on the gas content: we assume $k=1$.  The quantity  $\nu$ is the efficiency parameter of the star formation process that must be specified (see below). 

In the infall model, because of the interplay between gas accretion  and consumption, the SFR starts low, reaches a peak after a time approximately equal to $\tau$ and then declines. The functional form that could mimic this behavior is the time delayed exponentially declining law:

\begin{equation}
			\Psi(t) \propto \frac{t}{\tau}\exp\left(-\frac{t}{\tau}\right).
\end{equation}
		
\noindent 
The Schmidt law in eq. \ref{schmidtlaw} is therefore the link between gas accretion by infall and gas consumption  by star formation. 
		
As a whole, this kind of approach stands on a number of observational and theoretical arguments among which we recall: (i) the parameters $\nu$ and $\tau$ can be related to morphology \citep{Buzzoni2002} and to the presence of ongoing star formation activity inside observed galaxies \citep{Cassara_etal_2016}; (ii) the aforementioned quantities can be easily tuned in order to fit observational data, and also complex phenomena that would affect the rate of gas cooling, such as active galactic nuclei (AGN), can be empirically taken into account without going into detail \citep[see e.g.][]{Chiosi_etal_2017}.

The infall models we have described may include  many important physical phenomena, for instance gas heating by supernova explosions (both type Ia and type II), stellar winds, gas cooling by radiative emission, and the presence of galactic winds. See the study by \citet{Tantalo_etal_1998}  for all details on these topics. 

\subsection{Outline of the galaxy models}
 
The complexity of real globular clusters, galaxies and galaxy clusters and the history of their evolution are reduced here to ideal systems of which we know the current masses $M(t)$, $M_B(t)$, $M_D(t)$, $M_s(t)$, $M_g(t)$ together with the mass abundances of some important elements $X_i(t)$ (where $i$ stands for H, He, C, N, O, Mg, ... Fe) and total abundance of heavy elements $Z(t)$
\footnote{For more details on chemical enrichment, companion equations and chemical yields per stellar generation see \citet{Tantalo_etal_1998} }, and finally half-stellar mass (half-light) radius  $R_e(t)$, and  dark mass radius $R_D(t)$. At each time, the system contains a manifold of stellar populations of different metallicity and age which can be approximated by  single stellar populations (SSP) of mean metallicity $<Z(t)>$  and mean age $T(t)$ defined by the relation $T(t) = M_s(t) / <\Psi(t)> $ where $<\Psi(t)>$ is  the mean star formation rate in the interval $0 \div t$ (with $t$ the current age). This value of the age $T(t)$ will be used to infer the current luminosity associated to the stellar  content $M_{s}(t)$  (see below). 

The infall model of a galaxy must be completed  with the radii $R_e(t)$ and $R_D(t)$ that are necessary to calculate the velocity dispersion of the stellar component, and the gravitational potential for the onset of galactic winds. To this aim we shortly discuss a few items of interest here:

i) \textsf{The $M_D$-$M_L$ and $R_D$-$R_L$ relationships}. Following \citet{Bertin_etal_1992} and \citet{Saglia_etal_1992}, the  spatial distribution of the dark component with respect to the luminous one in dynamical models is such that  the mass and radius of the dark component ($M_D$, $R_D$) are related to the luminous ones ($M_L$,  $R_L$) by 
 
 \begin{equation}
  \frac {M_L(t)(t)}{M_D(t)} \geq \frac{1}{2\pi} \frac{R_L(t)}{R_D(t)}\left[1 + 1.37 \frac{R_L(t)}{R_D(t)}\right]
  \label{BerSag}
 \end{equation}
\noindent
 where we can pose  $M_L(t) \simeq  M_s(t)$,  $ M_D(t) = f_c M_B(t)  \geq f_c M_s(t)$  and $R_L(t) \simeq 2 R_e(t)$. Therefore knowing 
 $M_D(t)$, $M_s(t)$, and $R_e(t)$, we can get an estimate of $R_D(t)$ to be used in the calculation of the total gravitational potential. According \citet{Bertin_etal_1992} and \citet{Saglia_etal_1992} typical values are $M_L/M_D \simeq 0.2$  and $R_L/R_D \simeq 0.2$. Consequently within $R_e$ the mass of dark matter is small with respect to the stellar mass and can be neglected. Furthermore, the binding gravitational energy of the gas and stars is given by
\begin{equation}
{ \Omega_{j}(t)=-{\alpha}_{L} G  \frac{M_{j}(t) M_{L}(t)}{R_{L}(t) } - 
G  \frac{M_{j}(t) M_{D}} {R_{L}(t) } \Omega'_{LD}  }   
\label{gas_pot}
\end{equation}
\noindent
 {where $j$ stands for $g$ (gas)  or $s$ (stars)}, and $\alpha_L$ is a numerical factor $= 0.5$, and finally the term

\begin{equation}
{ \Omega'_{LD}= \frac{1}{2\pi} (\frac{R_{L}(t)}{R_{D}}) [1 + 1.37
            ( \frac{R_{L}(t)}{ R_{D} } )] }  
\label{dark_pot}
\end{equation}
\noindent
is the contribution to the gravitational energy given by the presence of dark matter. With the assumed ratios ${ M_{L}/M_{D}}$ and the above replacements  of $M_L$,  $R_L$ and $R_D$,
the  term ${ \Omega'_{LD}}$ is about 0.04. Therefore in the evaluation of the velocity dispersion of the stellar component via the VT the effect of DM can be neglected. 
 
ii) \textsf{Velocity dispersion}. The velocity dispersion of an object with $M_D(t)$, $M_s(t)$, and radius $R_e(t)$ is derived from the scalar VT: at each time an object is supposed to be very close to the condition of mechanical equilibrium and hence to satisfy the relation 
 
 \begin{equation}
 \sigma_s(t)   = \sqrt{ \frac{G} {k_v} \frac {M_s(t)}{ R_e(t)} }
 \label{eqvirsig}
\end{equation}

(iii) \textsf{The $R_e(M_s)$ relation}. The mass-radius relation (MRR) suited to our models is the empirical law {proposed by} \citet{Fan_etal_2010} in the context of the $\Lambda$-CDM cosmology. 
The expression is:

\begin{equation}
R_{e}=0.9 \left(\frac{S_S(n_S)}{0.34} \right) \left(\frac{25}{m}\right) \left( \frac{1.5}{f_\sigma} \right)^2 
\left( \frac{M_{D}}{10^{12}  M_\odot} \right)^{1/3} \frac{4}{(1+z_{f})}.
\label{mr3}
\end{equation}

\noindent
where $M_{D}$, $M_s$, and $R_{e}$ have their usual meaning; $R_e$ is in kpc; $z_f$ the redshift at which the collapse took place;   $S_S(n_S)$ indicates the shape of the baryonic component that in turn is related to the S\'ersic brightness profile  from which $R_e$ is derived; $n_S$  is the S\'ersic index;  
$f_\sigma$ is the three dimensional stellar velocity dispersion as a function of the DM velocity dispersion, $\sigma_s=f_\sigma \sigma_{D}$; and finally $m$ is the ratio $M_{D}/M_s$. 
We adopt here $S_S(n_S)=0.34$ and $f_{\sigma}=1$.  For more details see \citet{Fan_etal_2010,Chiosi_etal_2020} and references therein. 
The most important parameter of eq.(\ref{mr3}) is the ratio $m= M_{D}/M_s$ that is shortly discussed below.

The MRR of eq. \ref{mr3} is the locus of galaxy models on the MR-plane, the formation of which occurred at redshift $z_f$. It represents the position of model galaxies for different sources \citep{Chiosi_Carraro_2002, Merlin_etal_2012, Vogelsbergeretal2014}, however it does not correspond to the real MRR observed for objects from GCs to ETGs and GCGs because cosmological effects are also present \citep[the subject has been thoroughly discussed by ][  to whom the reader should refer for all details]{Chiosi_etal_2020}.

(iv) \textsf{The $M_{D}/M_s$ ratio}. Basing on the Illustris-1 data \citet{Chiosi_etal_2020}  have investigated how this ratio varies in the mass interval  $10^{8.5} <  M_{D} <10^{13.5}$ (masses are in $M_\odot$) and from $z=0$ to $z=4$ and proposed the  following relation:

\begin{equation}
 m \equiv \frac{M_D}{M_s} = (-0.223 z_f + 0.375) \log M_D + (3.138 z_f -3.430) \, .
\end{equation}
\label{ratio_MD_MS}
 
\noindent 
In the present study, however, we follow a different strategy that at each time step tightly correlates the mass in stars $M_s(t)$ to the total baryonic mass $M_B(t)$ and {the total} mass of dark matter $M_D(t)$. At each time we have $M_D(t) = f_c\, M_B(t)$ where $f_c$ is the cosmic proportion ($f_c \simeq 6$). The mass in stars $M_s(t)$ is determined  by the efficiency of star formation and in any case it is a fraction of the current baryonic mass. Therefore the parameter $m$ is given by the relation: 

\begin{equation}
 \log m(t) = \log \frac{M_D(t)}{M_s(t)} \,.
\label{ratio_md_ms}
\end{equation}

At the beginning of {a galaxy history} the ratio $m$ is very large and then declines tending to the limit value $f_c\simeq 6$, if the total baryonic mass is eventually turned into stars.  Examples of the time behaviour of the ratio  $m$ will be shown when presenting our model galaxies in some detail.

(v) \textsf{The star formation rate}. Thanks to  the short time scale of the energy input from massive stars (a few million years), compared to the mass accretion time scale by infall (from hundred to thousand million years) the galaxy is supposed not to differ from an equilibrium state so that the \citet{Talbot_Arnett_1975} formalism can be applied. \citet{Chiosi_1980} and \citet{Chiosi_Matteucci_1980} adapted the SFR of \citet{Talbot_Arnett_1975} to model disk galaxies in which the surface mass density of stars, gas and total baryonic mass are used and a suitable radial distance $\tilde r$ is introduced. 

We have adapted their formalism to our case (in which spherical symmetry is implicitly assumed), 

\begin{equation}
	\frac{d M_s (r,t)}{dt}  = - \frac{d M_g (r,t)}{dt} =
	\tilde{\nu} \left[\frac{M(r,t) M_g(r,t) } {M (\tilde{r}, t)}  \right]^{\kappa -1} M_g (r, t) 
	\label{sfr_1}
\end{equation}
where $M_g(r,t)$ and $M_s(r,t)$ are the mean mass densities of gas and stars within the generic sphere of radius $r$ at the time $t$, respectively. $M(\tilde {r}, t)$ is the total mass density (gas and stars)  within a  particular  radial distance from the galaxy center, and finally $\tilde  \nu$ is a parameter measuring the efficiency of star formation. The radius $\tilde {r}$ is a suitable radial scale  controlling star formation. In the Larson's view they might be associated to the radial distance at which the central spheroidal component exerts its tidal effect on the residual external gas. As a consequence of it, at any time the SFR is significantly inhibited at distances $r > \tilde{r}$. Since our models do not include any geometrical description, but deal a galaxy as  a point-mass entity whose mass varies with time, we drop the radial dependence of the SFR and the rate of star formation is simply reduced to:

\begin{equation}
	\frac{d M_s (t)}{dt}  = - \frac{d M_g (t)}{dt} =
	\tilde{\nu} \left[\frac{M(t) M_g(t) } {M(t)}  \right]^{\kappa -1} M_g (t) =  \tilde{\nu} M_g(t)^{\kappa}
	\label{sfr_2}
\end{equation}

Since in all infall models $M_g (t)$ increases by infall and decreases by star formation, the SFR starts low, reaches a peak after a time approximately equal to $\tau$ and then  declines.
By varying $\tau$ (time scale of the galaxy formation process) one can recover all types of star formation indicated by observational data going from GCs to LTGs and ETGs.
The infall scheme and companion SFR have been widely used in many studies on the subject  of galactic chemical evolution \citep[e.g.][for a  review and references]{Matteucci2016}. The infall galaxy model is very flexible and can be adapted to a wide range of astrophysical problems. Suffice it to recall that it has been adopted by \citet{Bressan_etal1994} to model the spectro-photometric evolution of ETGs reduced to point mass objects, extended by \citet{Tantaloetal1998} to the case of spherical systems  made of BM and DM mimicking ETGs, adapted by  \citet{Portinari2000} to include  radial flows of gas in disk galaxies, and recently used by \citet{Chiosi_etal_2017} to study the cosmic star formation rate and by \citet{Sciarrattaetal2019} to investigate the color-magnitude diagram of galaxies in general.

(vi) \textsf{The SF efficiency $\tilde{\nu}$}. In most galaxy models of this kind the specific efficiency of star formation 
$\tilde{\nu}$ is an external  free parameter to be adjusted according to the case under investigation. In this paper we follow a different strategy and derive $\tilde{\nu}$ from other properties of the models.  Starting from the idea put forward  by \citet{Brosche1970,Brosche1973} that the efficiency of star formation is driven by the velocity dispersion, we suppose that  $\tilde \nu$ can be written as:

\begin{equation}
\tilde{\nu} = \nu_0 \left[ \frac {\sigma_t}{ \sigma_s } \times \frac{\sigma_T}{\sigma_s} \right]^{0.5} 
\label{nu_eff}
\end{equation}

\noindent
where $\sigma_s$, $\sigma_t$, and $\sigma_T$ are the velocity dispersion  calculated using only the stellar component and  the total mass, both  measured at current time $t$ and present day age $T$. The factor $\nu_0$ depends on the choice made for $\kappa$ and secures the correct dimensions to $\tilde \nu$. For $kappa=1$, $\tilde \nu \equiv 1/t$. Finally, the harmonic mean between two different normalizations is meant to somehow cope with the uncertainty affecting the whole procedure. Since the stellar mass $M_s$ and radius $R_e$  grow with time, the efficiency is large at young ages and decreases with time toward the limit value of $\nu \simeq 1$.   

(vii) \textsf{Luminosity and specific intensity from mean SSPs}.  In order to calculate the B and V luminosities and the associated specific intensities  $I_{eB}$ and $I_{eV}$ of the stellar content of galaxy models in the course of their evolution,  we make use of the SSPs with the  \cite{Salpeter1955} IMF (slope in number x=-2.35, lower mass $M_l = 0.1 M_\odot$, upper mass $M_u = 100 M_\odot$, total SSP mass $M_{ssp} = 5.82 M_\odot$,  metallicity from  $Z=0.0004$ to $Z=0.04$,  6 values in total, and age from 10 Myr to 14 Gyrs) of the library by \citet{Bertellietal2008,Bertellietal2009,Tantalo_2005_due}. The absolute $M_B$ and $M_V$ magnitudes can be plotted against the logarithm of the age in years, and for each pass-band the mean age-magnitude relation is derived. Owing to the nearly linear behavior of each relationship, a linear fit is suited to get the relation between the mean absolute magnitude and the age $t$. These are given by: 

\begin{eqnarray}
\label{mbmv}
  M_B &=& 2.361 \log t -17.841    \label{mean_mag_age1}\\
  M_V &=& 1.975 \log t -14.886    \label{mean_mag_age2} .
\end{eqnarray}

\begin{figure}
 \centering
 {\includegraphics[width=8.0cm, height=8.0cm]{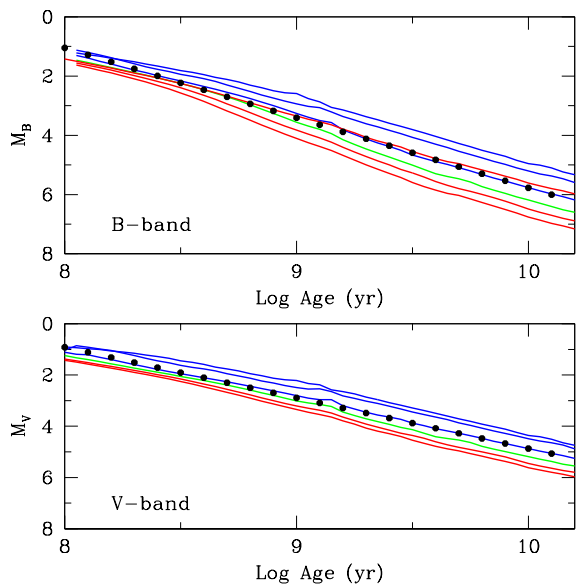} }
 \caption{The $M_B$ (top) and $M_V$ (bottom) magnitudes versus age relationships for SSPs of different metallicity according to the color code. From the top to the bottom the metallicity is  Z=0.0001, 0.001, 0.010, 0.019, 0.040, and 0.070.  The black dotted lines  are the mean values of $M_B$ and $M_V$  over the metallicity.}
\label{fig_15}
\end{figure}

\noindent
The age is expressed in years. The B and V magnitudes of the original SSPs with different metallicity are shown in Fig. \ref{fig_15} together with the metallicity averaged SSP (full dots).  The mean values of the magnitudes are meant to mimic the mixture of chemical compositions in a galaxy. At each time we know the {total mass made by stars} of different age and chemical composition. In practice we assume that this complex situation can be reduced to {a single} SSP of  the same mass, mean chemical composition (metallicity) and mean age T. The mean age is evaluated from the relation  $T(t) = M_s(t) / <\Psi(t)> $ where $<\Psi(t)>$ is  the mean SFR in the interval $0 \div t$ (with $t$ the current age). Using the mean age $T(t)$, from eqs.(\ref{mean_mag_age1}) and (\ref{mean_mag_age2}) we derive the B/V magnitudes (the luminosities) per unit mass of the SSP and then re-scale them to the mass $M_s$ of the galaxy.

(viii) \textsf{Solution of the basic equation eq.(\ref{eqbeta})}. At each time step of the evolutionary history of a model galaxy, known the star mass $M_s$, the radius $R_e$, the velocity dispersion $\sigma_s$, the luminosities $L_B$ and $L_V$ (in solar units) and the specific intensities $I_{eB}$ and $I_{eV}$, the equation system eq.(\ref{eqbeta}) is solved deriving $\beta$ and $L'_0$ at each time step. These are the two physical quantities that in our view drive the distribution of galaxies in the space of the physical parameters $L$, $R_e$, $\sigma$, and $I_e$, and determine the observed FP. 

\textsf{Final remarks}. The model age refers to the galaxy rest-frame and goes from $T_g=0$ at redshift $z_f$, when the galaxy is supposed to form, to $T_g=T_G$  at $z=0$ (present time). The corresponding ages of the Universe $T_U(z)$ are $T_U(z_f)$ and $T_U(0)$. For the $\Lambda$CDM cosmology with $H_0$ =71 km/s/Mpc, $\Omega_\Lambda=0.71$, $\Omega_m =0.23$, $\Omega_\Lambda=0.73$, $\Omega_{mD}/\Omega_{mB} \simeq 6$, we obtain $T_U(z_f)=0.484$ Gyr for $z_f=10$ and $T_U(0)=13.67$ Gyr and $T_G = 13.187$ Gyr. Whenever needed we will pass from one {to the other.} In order to minimize the number of free parameters in each model, we assume that all galaxies are born at the same redshift $z_f=10$; the collapse time scale of $\tau=1$ Gyr for all galaxies; the Salpeter initial mass function (in number) with a slope x=-2.35 and a  fraction of stars more massive than $1\, M_\odot$  equal to $\zeta=0.30$, and absence of galactic winds. However, a few cases will be shown for  different values of $\tau$, different values of $z_f$, and in presence of galactic winds. 

\subsection{Model results}\label{model_results}

In this section we discuss the galaxy models obtained with the above prescription for the infall scheme and star formation in particular. First, we  present the reference case with $\tau =1$ for all galaxy masses and the prescription for $\tilde \nu$ given by eq. (\ref{nu_eff}) together with the corresponding case $\tilde{\nu}= 1$ (we refer to these latter models as the reference case). Then we discuss some cases in which the effect of galactic winds energized by supernova explosions (both Type Ia and Type II) are taken into account. Table \ref{Tab_3} lists the models we have considered and {presents} some characteristic features at the last stage with active star formation: this is either the present age for the models without galactic wind or at the onset of galactic wind. In the following, we mainly present and discuss the models without galactic winds, limiting the discussion of those with galactic winds to some general remarks.

   \begin{table*}
    \begin{center}
    \caption{Tables of galaxy models. $M_B(t_G)$ in solar units is the present-day baryonic mass. Age is either the galaxy age at the present time or the age at the onset of galactic winds (ages in Gyrs).  $M_g$ and $M_s$ are the gas and stellar masses in solar units at the indicated age. $Z_g$ and $<Z_g>$   are the metallicity at the indicated age and the mean metallicity reached by the gas. SFR is the star formation rate in solar masses per year at the indicated age. Finally, $\Omega_g$ and $E_g$ are the gravitational energy and thermal energy of the gas at the onset of galactic winds. All energies are in units of $10^{30}$ ergs. In the case of models without galactic winds $E_g$ is not given.} 
    \label{Tab_3}
    \begin{tabular}{|c c c  c c c  c c c|}
     \hline
$M_B(t_G)$ &  Age    &  $  M_g $  &  $M_s$   &   $Z_g$  &$ <Z_g>$ &    SFR     & $\Omega_g$&  $E_g$   \\
\hline
\multicolumn{9}{|c|}{No Galactic Winds}\\
\hline
1e6        &  13.19  & 0.35E-02  & 0.94E+00  &   0.109  &   0.038 &  3.46E-06  & 1.98E-04  &   \\
1e8        &  13.19  & 0.35E-02  & 0.94E+00  &   0.109  &   0.038 &  3.46E-04  & 1.98E+00  &   \\
1e10       &  13.19  & 0.35E-02  & 0.94E+00  &   0.109  &   0.038 &  3.46E-02  & 1.91E+04  &   \\
1e12       &  13.19  & 0.35E-02  & 0.94E+00  &   0.109  &   0.038 &  3.46E+00  & 4.02E+07  &   \\
5e12       &  13.19  & 0.35E-02  & 0.94E+00  &   0.109  &   0.038 &  1.73E+01  & 2.43E+08  &   \\
\hline
\multicolumn{9}{|c|}{Galactic Winds}\\
\hline
1e6        &  13.19  & 0.63E-03  &  0.99E+00 &   0.038  &   0.012 &  6.28E-07  &  3.60E-05 & 2.2E-05  \\
1e8        &   7.46  & 0.96E-02  &  0.98E+00 &   0.056  &   0.017 &  9.55E-04  &  5.46E+00 & 5.7E+00  \\
1e10       &   5.75  & 0.44E-01  &  0.92E+00 &   0.089  &   0.033 &  4.35E-01  &  2.39E+05 & 2.5E+05  \\
1e12       &   5.25  & 0.74E-01  &  0.88E+00 &   0.109  &   0.045 &  7.35E+01  &  8.50E+08 & 8.8E+08  \\
\hline
\end{tabular}
\end{center}
\end{table*}
 
\begin{figure}
 \centering
 {\includegraphics[width=8.0cm, height=8.0cm]{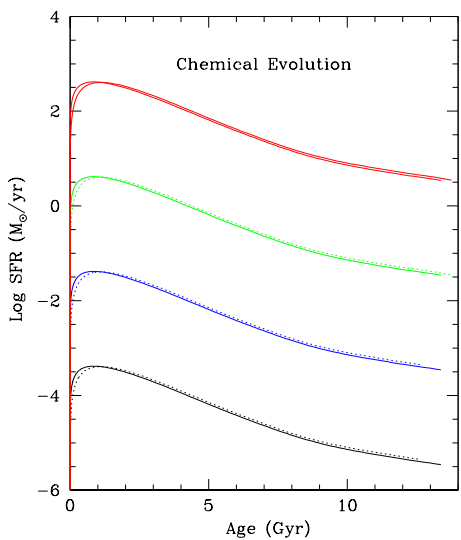} }
 \caption{The SFR histories of models with $M_{B}(T_{G})$ equal to  $10^6$, $10^8$, $10^{10}$, and $10^{12}$ $M_{\odot}$ (black, blue, green, and red in the order) and variable $\tilde{\nu}$ (solid lines) and $\tilde{\nu} =1$ (dashed lines, the reference case). }
\label{fig_16} 
\end{figure}

\begin{figure}
 \centering
 {\includegraphics[width=8.0cm, height=8.0cm]{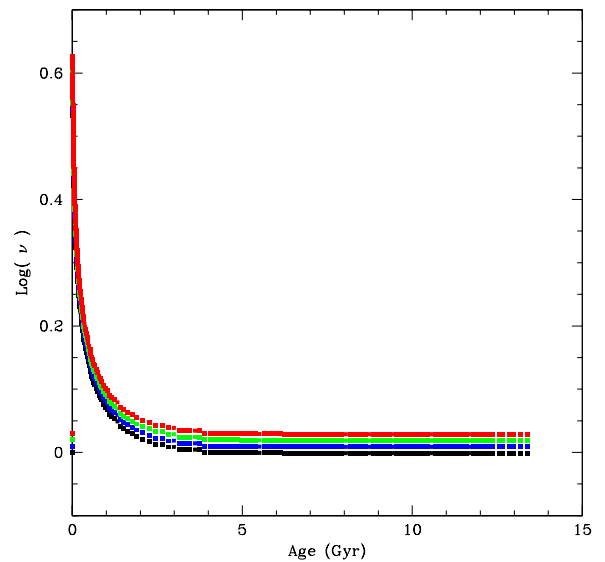} }
 \caption{The temporal variation of the SF efficiency of the galaxy models with with $M_{B}(T_{G})$ equal to $10^6$, $10^8$, $10^{10}$, and $10^{12}$ $M_{\odot}$ (black, blue, green, and red in the order). The efficiency is the same for all models and goes from 4.04  to 1. Plotting the data $\log(\tilde{})$ of each case has been shifted by 0.01 with respect to the other ones.  }
\label{fig_17} 
\end{figure}

\textsf{SFR and SF efficiency}.  In Fig. \ref{fig_16} we show the history of star formation in $M_{\odot}/yr$ of galaxies with $M_{B}(T_{G})$ equal to $10^6$, $10^8$, $10^{10}$, and $10^{12}$ $M_{\odot}/yr$ (black, blue, green, and red in the order) and variable $\tilde{\nu} $ (solid lines) and $\tilde{\nu} =1$ (dashed lines, the reference case). As expected, the SFR starts { small, reaches} a peak value and  then declines  to low values even though it never extinguishes. The peak value is at an age nearly equal to the infall time scale. Models with variable efficiency do not differ from their corresponding reference case with constant $\tilde\nu=1$. The reason for it resides in the  value of $\tau$. The point will be clear discussing the case in which $\tau$ is let change with the galaxy mass.  The SFR efficiency $\tilde{\nu}$ is a dimensionless quantity and therefore is the same for all galaxies; it varies with time from the initial top value 4.04 to 1 as shown in Fig.\ref{fig_17}. 
 
The advantage of our choice for the SF efficiency $\tilde{\nu}$ {is that} this important physical quantity is no longer a free parameter. It is indeed deeply driven by the galaxy mass building process and the time scale associated to it. With our choice for $\tau$, the SF efficiency very quickly reaches its asymptotic value (within about $2 \times \tau$). If $\tau$  is increased the time scale over which $\tilde{\nu}$ goes to the asymptotic value gets accordingly longer.
 
\begin{figure}
 \centering
 {\includegraphics[width=8.0cm, height=8.0cm]{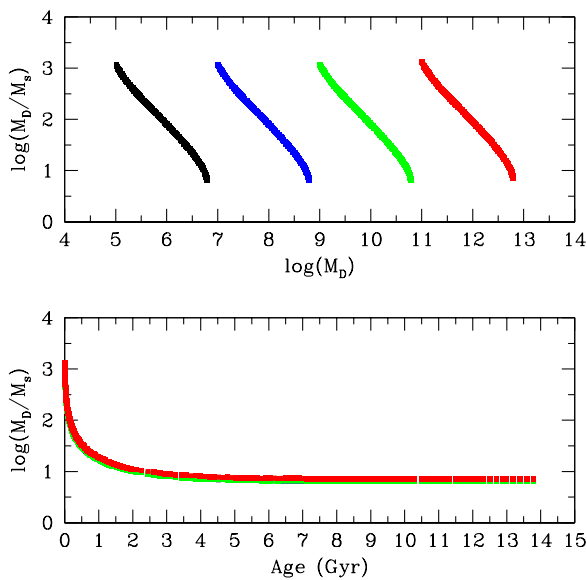} }
 \caption{The ratio $m= M_D/M_s$ as a function  of the  mass of Dark Matter $M_D$ (top panel) and age (bottom panel) for our model galaxies. Masses are in $M_\odot$ and ages in Gyrs. The model galaxies are in different colors and ranked according to their total baryonic mass $M_B$ reached at the present time (black: $10^6$; blue: $10^8$; green: $10^{10}$, and red: $10^{12}$). In the bottom panel all the lines overlap each other. }
\label{fig_18} 
\end{figure}

\textsf{The ratio $M_D/M_s$ and the radius $R_e$}. The stellar radius $R_e$ depends on the Dark Mass to  stellar mass ratio $M_D/M_s$. As already  explained this ratio is determined at each time  from the current value of the stellar mass built up by star formation, the current mass of baryonic mass $M_B(t)$ and the current mass of Dark Matter associated to it given by $M_D(t) = f_c M_B(t)$. The ratio $M_D/M_s$ is shown in Fig.\ref{fig_18} as a function of the mass $M_D$ (top panel) and age in Gyr (bottom panel). In each galaxy the ratio starts very high and, as time increases, tends to the limit value $f_c \simeq 6$ as the whole baryonic gas mass is  turned into stars by star formation. The general behaviour of $M_D/M_s$ as a function of $M_D$ and age is the same to the point that in the  bottom panel all the curves overlap. Also in this case the ratio $M_D/M_s$ is not an external parameter, but it is determined in a self consistent way by the internal properties of the models. 
 
\begin{figure}
 \centering
 {\includegraphics[width=8.0cm, height=8.0cm]{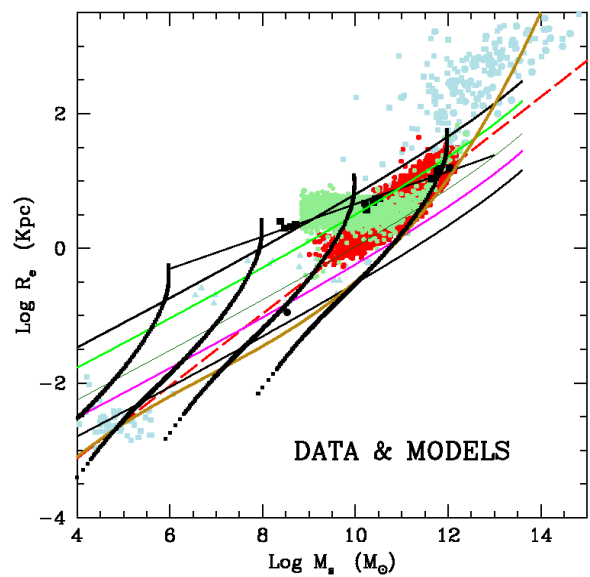} }
 \caption{The mass-radius relations (MRRs) of our model galaxies labelled by their present day total baryonic mass $M_{B}(T_{G})$ equal to $10^6$,  $10^8$, $10^{10}$, and $10^{12}$ $M_{\odot}$ from left to right. Each line made by filled black squares represents the whole evolutionary history of both $M_s$ and $R_e$ both increasing with time (the top is the present). These models are compared both with observational  and theoretical data from different  sources: (i)  the observational data of \citet{Burstein_etal_1997} from GCs to GCGs (powder-blue small dots) and the ETGs by \citet{Bernardi_etal_2010} (red small dots); (ii) the Illustris-1 galaxies (light green small dots); (iii) the  low initial density models (blue squares and their best fit) and the high initial density ones (red squares and their best-fit) by  \citet{Chiosi_Carraro_2002}; (iv) the early hierarchical models by \citet{Merlin_etal_2012} (black squares and their best-fit); (v)  the   \citet{Fan_etal_2010} MRRs for different values of the formation redshift $z_f =0$ (top), 1, 5 10, and 20 (bottom); (v) finally and the MRR  by \citet{Chiosi_etal_2020} (dark golden line). See the text for more details.  }
\label{fig_19} 
\end{figure}

With the aid of the $m$-ratio and the MRR of eq. (\ref{mr3}) we  derive the radius $R_e$ of the stellar component $M_s$ and build the mass-radius relationship (MRR) of our model galaxies shown in Fig. \ref{fig_19} both along their evolutionary history (the black line drawn by filled squares, one for each time step, where  the present time is at the top and the initial stage at the bottom). Each curve corresponds to a model with a different final total baryonic mass $M_B$, namely  $10^6$ $M_\odot$,  $10^8$ $M_\odot$, $10^{10}$ $M_\odot$,  $10^{12}$ $M_\odot$ from left to right. Although the MRRs of the models are in fair agreement with the bulk of observational data and other theoretical MRRs, the closer inspection of the issue reveals that our theoretical radii are likely overestimated by a factor that is difficult to assess. Our best estimate is about a $\Delta log R_e \simeq -0.6$ to $-0.8$. The mean radii should be a factor 4 to 6 smaller. There are many possible causes for this disagreement: first of all, in addition to the $m$-ratio in the term $(M_D)^{1/3}$ eq. (\ref{mr3}) contains other terms each of which is affected by some uncertainty. The terms in question are the ratio ${}S_S(n_S)/{0.34}$, the ratio $( {1.5}{f_\sigma})^2$, and finally  the ratio  $({25}/{m})$. The first two are simply assumed to be equal to one, while the last one contains the ratio $m$ and deserves some remarks. It is clear that it has been introduced as an adjustment factor based on some estimates of the $m$-ratio derived from current theoretical N-Body Smoothed Particle Hydrodynamic (NBTSPH) simulations of galaxy formation in which only a small fraction of the available gas was used to form stars \citep[e.g. see for instance][]{Chiosi_Carraro_2002}, which explains the factor 25. The present infall models have a different behaviour because nearly all the gas is used up to form stars and the limit value of the $m$-ratio {is about $f_c \simeq 6$}. This implies that the above  adjustment factor should become $({f_c}/{m})$,  and consequently a reduction of the estimated radius by a factor of about 4 to 6.  
However, instead of forcing the radius to strictly agree with the data, thus introducing some ad hoc adjustments, we keep the radii as they are but also keep in mind that in reality they could be 4 to 6 times smaller than estimated. This would immediately affect our evaluation  of the specific intensity $I_e = L/(2 \pi R_e^2)$ that could be a factor 16 to 36 higher than our straight evaluation (see below). 
 
\begin{figure}
 \centering
 {\includegraphics[width=8.0cm, height=8.0cm]{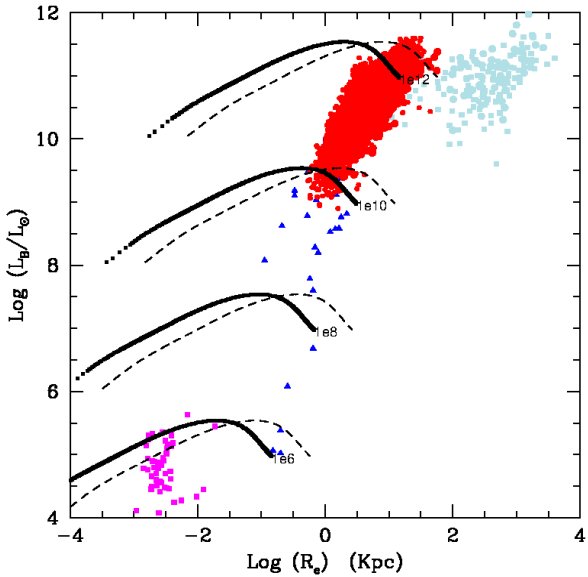} }
 \caption{The B-luminosity-radius relation (L$_{B}$RRs) of our model galaxies with $M_{B}(T_{G})$ equal to $10^6$,  $10^8$, $10^{10}$, and $10^{12}\, M_{\odot}$, from bottom  to top. For each mass we display two lines: the one with the original radii (dashed black line)  and the case with the radii decreased by a factor of 4 as explained in the text (line made  by filled black squares). The models are compared both with observational  data from \citet{Burstein_etal_1997} from GCs (magenta small dots) to Dwarf Galaxies (blue small dots), to ETGs (red small dots), GCGs (powder-blue small dots), and the ETGs by \citet{Bernardi_etal_2010} (red small dots, overlap the previous ones). The agreement for the smaller radii case is soon evident. See the text for more details. }
\label{fig_20} 
\end{figure}

\begin{figure}
 \centering
 {\includegraphics[width=8.0cm, height=8.0cm]{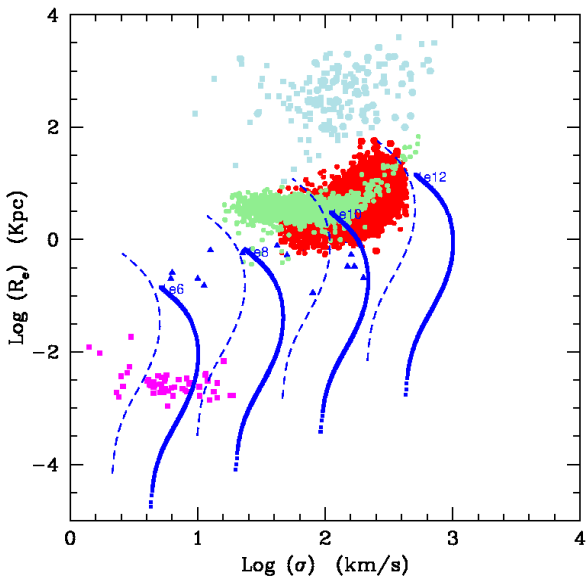} }
 \caption{The Re-$\sigma$ relation (RSR) of our model galaxies. In this figure the same models, observational data, color code, and symbols used in Fig.\ref{fig_20} are adopted. }
\label{fig_21} 
\end{figure}

\textsf{Other important relationships: the $L$ vs $R_e$ and the $R_e$ vs $\sigma$}. The  uncertainty on the radius affects also other important relationships such as the luminosity-radius relation (LRR) shown in Fig \ref{fig_20} and the radius velocity dispersion relation (RSR) displayed in Fig. \ref{fig_21}. The theoretical data are compared with the observational ones by \citet{Burstein_etal_1997} and \citet{Bernardi_etal_2010}, for these latter the mean colour (B-V)=0.85 has been applied to the $M_V$ magnitudes to  get the B-luminosity. In both cases  the best results are for radii reduced by a factor of 4.

\begin{figure}
 \centering
 {\includegraphics[width=8.0cm, height=8.0cm]{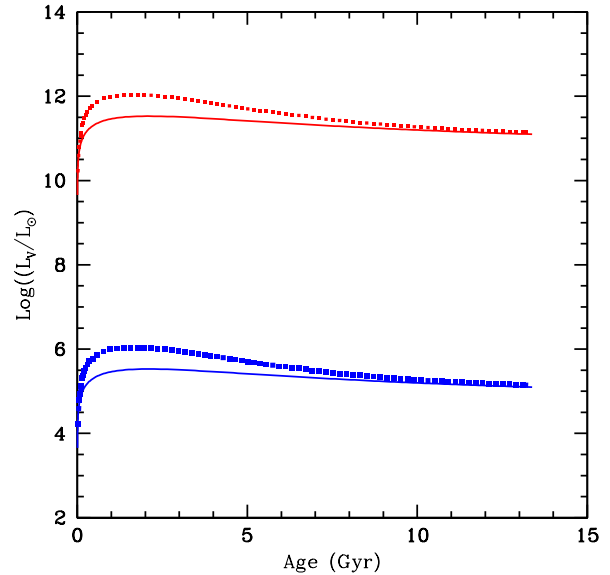} }
 \caption{ The luminosity $L_V$ versus age (in Gyr)  relation (LAR) of our model galaxies. The cases $M_{B}(T_{G})= 10^6$ (bottom) and   $10^{12}\, M_\odot$  (top) are shown. The solid lines are luminosities derived from the analytical relationships while the dotted lines are those from full population synthesis. }
\label{fig_22} 
\end{figure}

\textsf{Remarks on the luminosity}. Before proceeding further it is worth commenting on the luminosity of the model galaxies. As already explained, {for the sake of a quick assessment  of the model galaxies luminosity } in the B and V pass-bands, we have used suitable linear relationships between the absolute B and/or V  magnitudes of the Johnson system and the mean age $T$ based on SSPs of mean metallicity. One may argue that the luminosities derived in this way are much different from those evaluated by means of full population synthesis technique, i.e. by integrating the spectral energy distribution of SSPs over the star formation rate,  initial mass, and the metallicity range spanned by the stellar populations of galaxies at each time  
\citep[see][for all details]{Bressan_etal1994}. This is done "a posteriori", once the whole SFR(t), $M_s(t)$, and metallicity $Z(t)$ are  known. The results are shown in Fig.\ref{fig_22} for the $M_{B}(T_{G})= 10^6,  10^{12}\, M_\odot$, where the solid lines are the luminosity from the analytical relationships, and the dotted lines the luminosity from full population synthesis. The two luminosities {differ from each other by a maximum factor} of 3 back in the past when the SFR(t) was maximum (ages of about 1.5 Gyr), while they coincide in the less remote past (roughly past 5-6 Gyr). Therefore our approximation that nicely speeds up the model calculation is reasonable and leads to acceptable results. Our luminosities can be safely used for the present purposes.

\begin{figure}
 \centering
 {\includegraphics[width=8.0cm, height=8.0cm]{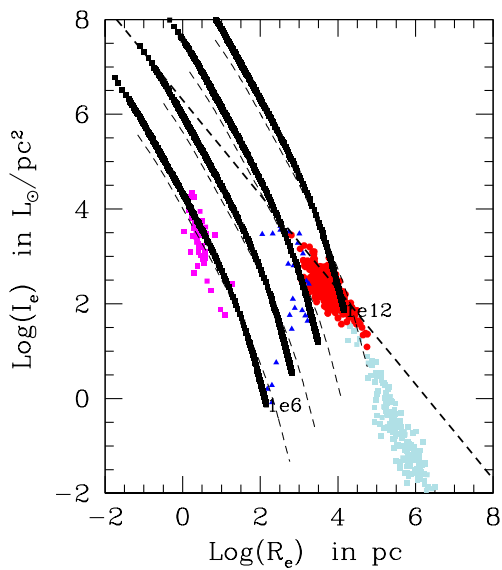} }
 \caption{The $I_e$-$R_e$ plane  of our model galaxies {compared} with the observational data of \citet{Burstein_etal_1997}. The color code of the data is the same as in previous figures. There are two groups of models: the  black thin dashed lines are models with original radii, while the thick lines made by filled black squares are those with the radius decreased by a factor of 4 and the specific intensity $I_e$ increased by a factor of 16. The galaxy mass is $M_{B}(T_{G})$ equal to $10^6$,  $10^8$, $10^{10}$, and $10^{12}\, M_{\odot}$, from left to right. Along each line the time runs from zero to present age from the top to the bottom. The formation redshift of all the models is $z_{for}$=10.}
\label{fig_23} 
\end{figure}

\textsf{The $I_e$ vs $R_e$ plane}. Together with the FP and the luminosity-velocity dispersion relation, otherwise known as Faber-Jackson (FJ) relation, the \IeRe\ plane is one of the most studied projection of the FP. The uncertainty on the radius $R_e$ (a factor of 4) reflects on $I_e$ as an increase of a factor of 16 at fixed stellar mass of the galaxy. The results for our models (with no galactic winds) are shown in Fig. \ref{fig_23} and are compared with the observational data of \citet{Burstein_etal_1997} from GCs to GCGs using the same color code as in Fig.\ref{fig_20}. The evolutionary sequences on display are for model galaxies with $M_{B}(T_{G})$ equal to $10^6$,  $10^8$, $10^{10}$, and $10^{12}\, M_{\odot}$, from left to right. For each mass we display two lines: the one with the original radii (dashed black line)  and the case with the radii decreased by a factor of 4 and the specific intensity $I_e$ increased by a factor of 16 as explained in the text (line made  by filled black squares). The time evolution goes from the top to the bottom of each line. The present day stage is the last bottom point of each line. Finally the thick dashed line is the border of the ZOE. Please note that no model at the present time falls in the ZOE, but all are below it. 

The present models cannot account for the data of GCs (as expected). Even if the model with $M_{B}(T_{G})=10^6\, M_\odot$ crosses the region of GCs it cannot reproduce these objects  because the present radius and specific intensity are too large and too low respectively. The ongoing star formation yields too luminous and too large objects that do not match with general properties of GCs. What would be needed are models in which star formation ceased and radius stopped growing  long ago (short initial episode followed by quiescence perhaps because of strong galactic wind), or to take into account the important transformations induced by the interaction with the Galaxy. 

Similar considerations can be applied to clusters of galaxies for which different type of models should be set up. To develop a suitable model for the formation and evolution of GCGs along the same lines we have followed for single galaxies is beyond the aims of this study and we leave the subject to a future investigation. 

\begin{figure}
 \centering
 {\includegraphics[width=4.0cm, height=8.0cm]{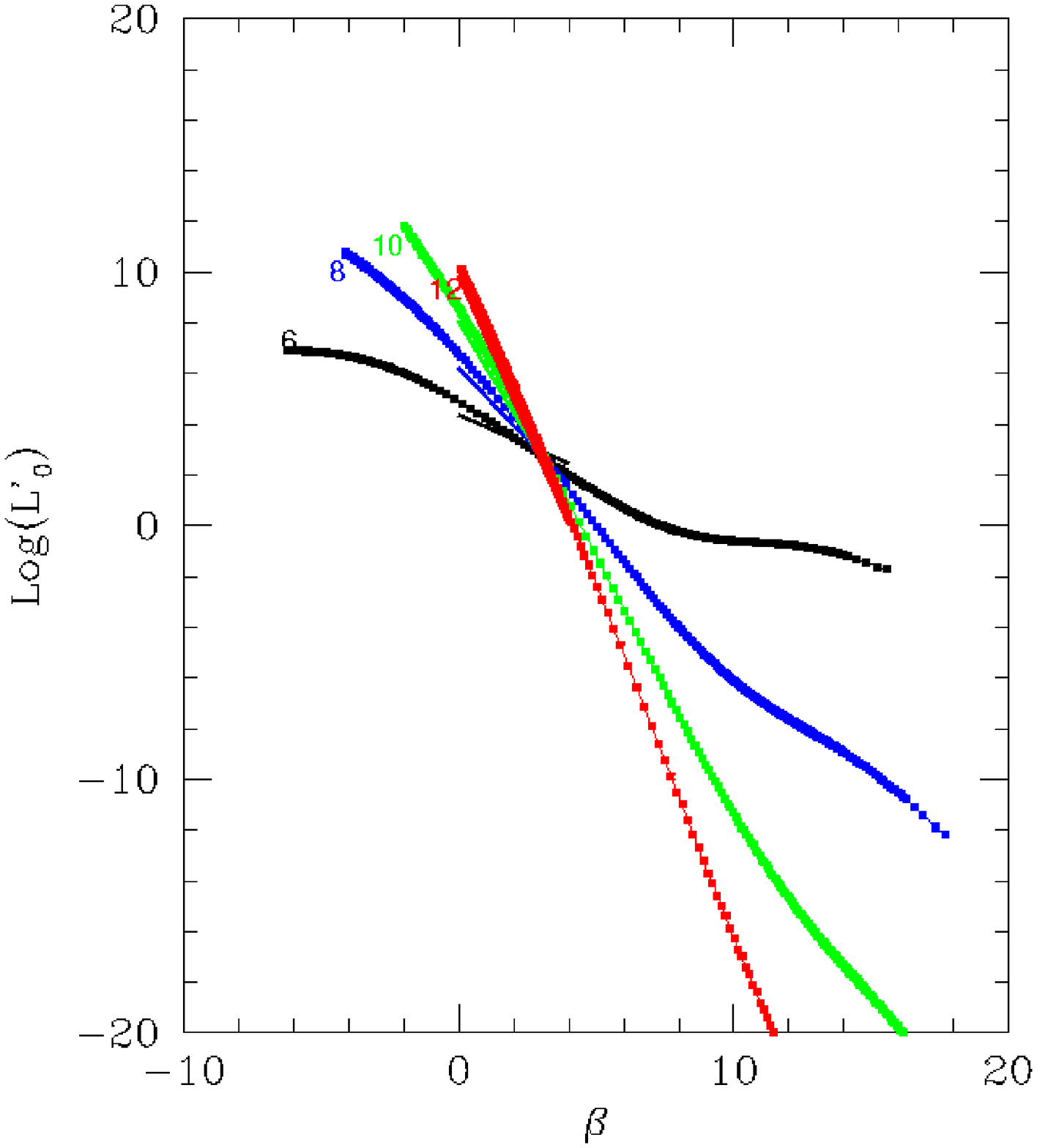} }
 {\includegraphics[width=4.0cm, height=8.0cm]{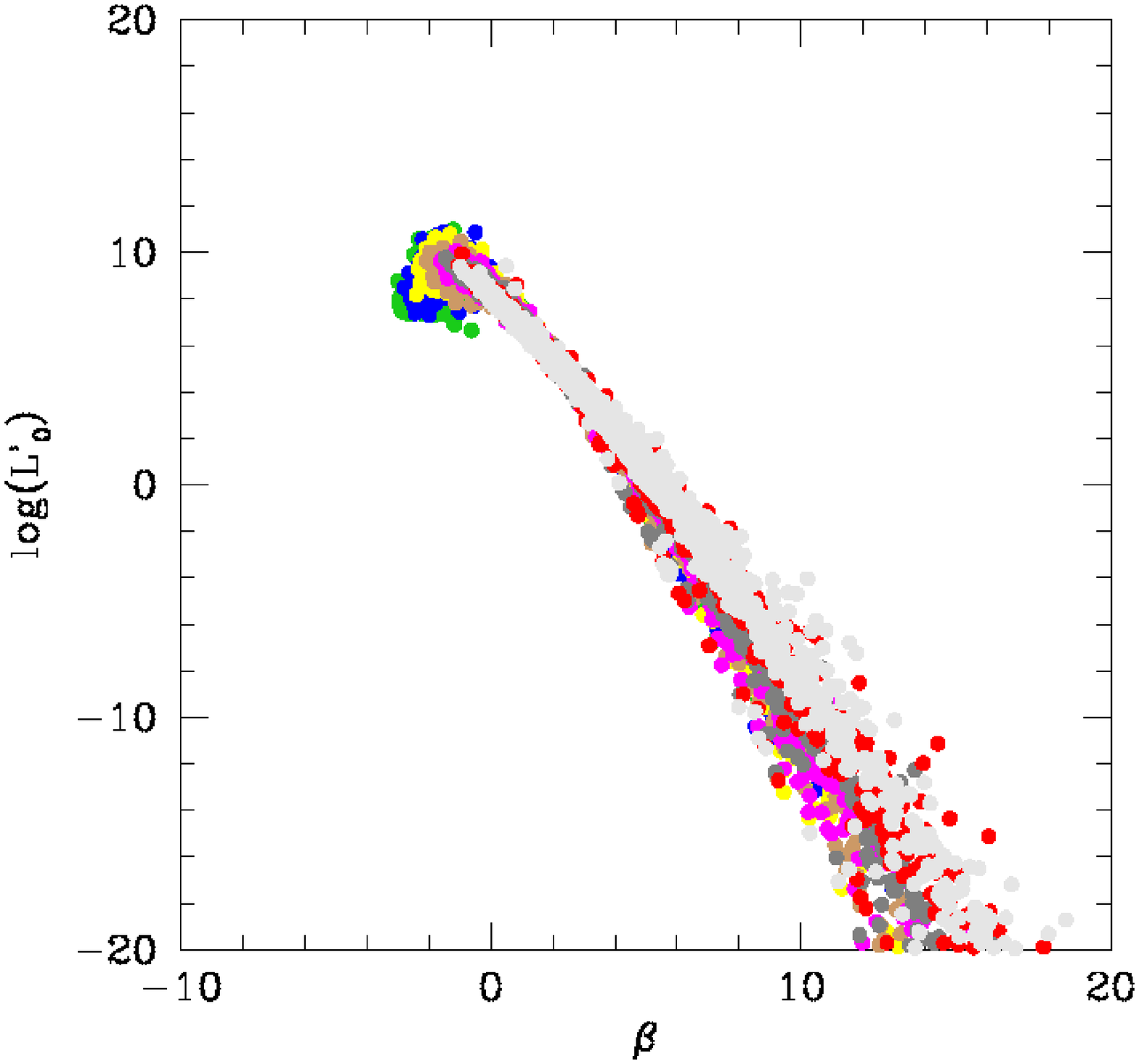} }
 \caption{Left panel: The $L'_0 - \beta$ relation of our models (left panel) evolved without galactic winds. All the relationships are nearly identical passing from models with large radii to those with smaller radii (by  a factor of 4), and the corresponding solutions of the equations for $\beta$ and $L'_0$, and finally changing only the photometric pass-band in use. The relationships seem to depend indeed only on the galaxy mass.
 Right panel: The $L'_0 - \beta$ relation for the artificial galaxies of Illustris-1. Each color corresponds to a different redshift epoch: green ($z=4$), blue ($z=3$), yellow ($z=2.2$), brown ($z=1.6$), magenta ($z=1.0$), dark gray ($z=0.6$), red ($z=0.2$) and light gray ($z=0$). }
\label{fig_24} 
\end{figure}

\textsf{The $\beta - L'_0$ space}. With the aid of the equations from eq.(\ref{eqsig}) to eq. (\ref{eqbeta}) we derive the exponent $\beta$ and proportionality factor $L'_0$ along the whole evolutionary sequence of our model galaxies evolved without galactic winds. In the left panel of  Fig. \ref{fig_24} we display all cases under consideration: models with large radii and models with smaller radii (the factor of 4) for the two photometric pass-bands in usage (B and V Johnson). Along each line time increases from the bottom to the top where the last stage at the present age is indicated by the mass label (total asymtpotic baryon mass). For each galaxy mass ($M_B(t_G)$) the results are nearly the same, all sequences overlap each other. See also the entries of Table \ref{Tab_4} containing the slope $\alpha$ and zero-point $\gamma$ of their linear best-fits. It turns out that the relationships in question depend only on the galaxy mass. Remarkably $\beta$ the exponent of the $L=L'_0 \sigma^\beta$ relation is positive during the early stages and negative afterwards. The luminosity first increases with $\sigma$ and then decreases with it afterwards. Finally note that  all curves cross each other at $\beta\simeq 3$ and $\log L'_0 \simeq 2.5$, values very close to the observed FJ relation. To confirm  this picture,  in the right panel of Fig. \ref{fig_24}  we plot the same relations for the Illustris-1 models grouped at different redshifts from $z=4$ to $z=0$. Now the situation is not the same as before, because in each group with the same redshift, mass and age vary from galaxy to galaxy. Furthermore not all masses are present at each redshift: samples at high redshifts, say $\geq 1,6$, are dominated by low mass objects (masses lower than $10^8 \, M_\odot$ are missing anyway because of the mass resolution), massive objects up to $10^{12}\, M_\odot$ are present at lower redshifts. However, the resulting distributions in the  $\beta - L'_0$ plane are much similar to that of the left panel, and remarkably there is also some evidence of the $\beta \simeq 3$ cross-point. This fact strongly supports the notion that infall models nicely mimic the numerical hierarchical simulations. 

\begin{table}
 \begin{center}
  \caption{The relationships between $L'_0$ and $\beta$  for the model galaxies evolved without galactic wind.  These relationships are the linear best fits of the curves shown in Fig.\ref{fig_24}. All these relationships are nearly identical passing from models with large radii to those with smaller radii (by  a factor of 4), and the corresponding solutions of the  equations for $\beta$ and $L'_0$, and finally changing only the photometric pass-band in use. The relationships seem to depend indeed only on the galaxy mass. }
  \label{Tab_4}
  \begin{tabular}{|c  |c |c |c  c|}
   \hline
  \multicolumn{5}{|c|}{$log L = \alpha * \beta  + \gamma$}\\
  \hline
  \multicolumn{1}{|c|}{}&
  \multicolumn{2}{|c|}{B-Band}&
  \multicolumn{2}{|c|}{V-band}\\
  \hline
  $M_B/M_\odot$  &  $\alpha$ & $\gamma$ &  $\alpha$ & $\gamma$\\
  \hline
  1e6            & -0.478    & 4.369     & -0.508    & 4.368  \\
  1e8            & -1.137    & 6.224     & -1.167    & 6.287   \\
  1e10           & -1.796    & 8.046     & -1.760    & 7.340   \\
  1e12           & -2.453    & 9.834     & -2.474    & 9.888   \\
  \hline
  \end{tabular}
 \end{center}
\end{table}

\begin{figure}
 \centering
 {\includegraphics[width=8.0cm, height=8.0cm]{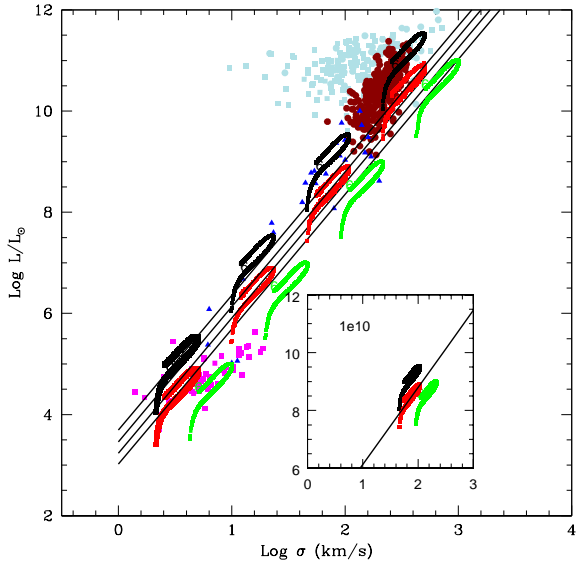} }
 \caption{The luminosity $L_B/L_\odot$ versus velocity dispersion $\sigma$ in (km/s) relation (LSR) of our model galaxies. For each mass (labelled by $M_{B}(t_G)$ as indicated ) three relations are shown: (1) the original models with no revision of the radii (lines made by filled black squares); (2) models whose luminosity is derived from {the $ L_B/L_\odot = L'_0 \sigma^\beta$ relation } with  the original $\beta$ and $L'_0$ (the curves made by red squares)  together with the linear fit limited to  the descending branch of each curve (the black solid lines); (3) models in which the radii have been revised  and new values of $\beta$ and $L'_0$ are calculated (the green curves). Note how in each case, the luminosity vs sigma relation bends  past the stage that roughly corresponds to the maximum stellar activity. From this stage the luminosity and velocity dispersion decrease (see the text for details). The insert shows the  case of the $M_{B}(t_G)=10^{10}\, M_\odot$ for the sake of better illustration. The models are compared with the data by \citet{Burstein_etal_1997} from GCs to GCGs (the same color code as in previous figures is used). } 
\label{fig_25} 
\end{figure}

The most important relation to look at and to examine in detail is the luminosity versus velocity dispersion. This is shown  in Fig.\ref{fig_25} which displays  the  $\log(L_B/L_\odot$) vs  $\log \sigma$ for the model galaxies and compare it with observational data of \citet{Burstein_etal_1997}. In the main panel we display three possible relationships: (i) the plain $L_B/L_\odot$ vs $\sigma$ of the models with their original luminosity and radii (the thick black curves). Along each curve the evolution starts at the bottom point of each line and proceeds to the final stage indicated by the label  $M_B(t_G)$ in solar units. Please note that during the galaxy lifetime the $L_B/L_\odot$ vs $\sigma$ relation bends over past a certain age toward lower luminosities and lower velocity dispersion. This  roughly happens  past the peak of star formation. While the luminosity decrease can be easily understood, the decrease in velocity dispersion of the stellar component needs some explanation. Stars during their lifetime can explode as Type II and Type I supernovae: in the first case a small remnant is left (neutron star or black hole), in the second one no remnant at all. They can also lose lots of mass by stellar winds. In any case the total mass in stars is expected to decrease and so does the velocity dispersion.  
(ii) The second case is the associated $ L_B/L_\odot = L'_0 \sigma^\beta$ relation in which the original $\beta$ and $L'_0$ are used (the red curves together with the linear fit limited to  the descending branch of each curve, (the red solid lines). 
(iii) Finally, the $ L_B/L_\odot = L'_0 \sigma^\beta$ relation, in which the correction on the radius has been applied and new values of $\beta$ and $L'_0$ are derived. It is worth recalling that the  $L'_0$ vs $\beta$ relation  remains unchanged. The results are shown by the green curves. The small insert in Fig. \ref{fig_25} shows the case of the $M_B(t_G)=10^{10}\, M_\odot$ in more detail for the sake of better understanding. Similar results are found for the V pass-band that are not shown here. Two important features are soon evident; first of all the relations $\log (L_B/L_\odot$) vs  $\log \sigma$, based on the model history past the star formation activity period  have a similar slope, but different zero point (that depends on the galaxy mass). The manifold of these relations provides a sort of natural width to the luminosity-sigma relationship. The mean slope of the manifold agrees with the current value of the observed FJ. Second, the theoretical relations marginally agree with the body of ETGs, a steeper slope at luminosities above $\log L_B/L_\odot \simeq 9$ would be more appropriate. 

The simplicity of the current models cannot lead to better results. 
A possible improvement could be given by allowing small secondary episodes of star formation. The argument is as follows. The luminosity is the product of the star mass times the flux per unit mass: {let us } call $L_o = f_o M_o$ the original luminosity and $L_n = f_n M_n$ the expected luminosity including some recent star forming activity; $L_n M_n$ is in turn made by $f_y M_y + f_o M_o$, where $f_y M_y$ is the contribution by the episodic stellar activity; it follows that $(f_y M_y + f_o M_o)/(f_o M_o) = \lambda = L_n/L_o$. Basing on the current observations one would expect $ \lambda  \simeq 2$ or so. Now we may also assume  $M_y << M_o$ so that the total stellar mass and velocity dispersion in turn remain  nearly constant. Indicating with $\theta = M_y/M_o$, one gets $f_y / f_o= \lambda - 1 / \theta \simeq 5 - 10$ which is not impossible according to current population theories for Single Stellar Populations.
(iii) In the theoretical models the exponent $\beta$ of the $L_B/L_\odot = L'_0 \sigma^\beta$ relationship (a generalization of the FJ) can be either positive or negative depending on the particular evolutionary stage of the galaxy. Therefore among the observational data both values of $\beta$ are to be expected without violating the  trend indicated by the FJ, i.e. that the luminosity of galaxies increases with the velocity dispersion, hence the mass of the galaxy. 

\textsf{Galactic Winds}. Long ago \citet{Larson1974} postulated that the present-day Color-Magnitude Relation (CMR) of ETGs could be the result of {galactic winds} powered by supernova explosions, thus initiating a long series of chemo-spectro-photometric models of elliptical galaxies standing on this idea  \citep[see for instance][and references]{Tantalo_etal_1998}.  In brief, gas is let escape from the galaxy and star formation is supposed to halt when the total thermal energy of the gas equates its gravitational binding energy. This idea has been extended including the effect of stellar winds in the thermal energy budget of the gas. It was also included in NBTSPH models of galaxies \citep[see][and references]{Merlin_etal_2012}.
 
The same scheme proposed by \citep[][]{Tantalo_etal_1998} is adopted here, however with minor modifications because of the much simpler present formalism. As already said,  the thermal energy of the gas is the sum of three contributions, namely type I and  II supernovae and stellar winds from massive stars: 

\begin{equation}
E_{th}(t) = E_{th}(t)_{SNI} + E_{th}(t)_{SNII} + E_{th}(t)_{W} 
\label{Eth_tot}
\end{equation}

\noindent
where each term has the generic expression 

\begin{equation}
E_{th}(t)_{j} = \int_{0}^{t} \epsilon_{j}(t-t') 
            R_{j}(t') M_{B}(t_{G}) dt' 
\label{Esni}
\end{equation}

\noindent
with j= SNI, SNII, W with obvious meaning of the symbols. 
The normalization factor $M_{B}(t_G)$ in the above equations is required to calculate the energy in physical units.
The time $t'$ is either the SN explosion time or the time of ejection of the stellar winds as appropriate. The functions $\epsilon_{SN}(t)$ and $\epsilon_{W}(t)$ are cooling laws governing the energy content of supernova remnants and stellar winds, respectively. Finally, star formation and chemical enrichment are halted, and the remaining gas content is supposed to be expelled out of the galaxy (winds) when the condition

\begin{equation}
E_{th}(t) \geq \Omega_{g}(t)
\label{eth_omg}
\end{equation}

\noindent
is verified. For all other details concerning the above rates, the evolution of SN remnants and stellar winds and how much of the initial energy budget is shared with the gas to energize it, and finally the expression for the gravitational energy of the gas in presence of baryonic and dark mass and  their space distribution in a galaxy see \citet{Tantalo_etal_1998}.

A small sample of models with galactic winds are calculated and their main features are summarized in Table \ref{Tab_3}. It is worth noting that the onset of galactic winds occurs at younger and younger ages as the galaxy mass increases.  Thanks to it, these models obey the constraint imposed by observational data on chemical elements like Carbon (C), Oxygen (O),  Magnesium (Mg), also known as $\alpha$-elements,    and Iron (Fe) and their ratios $[\alpha / Fe]$: the high mass galaxies are more $\alpha$-enhanced ($[\alpha / Fe] > 0 $) than the low-mass ones ($[\alpha / Fe] \leq 0 $). This fact cannot be easily reconciled  with other properties of the same objects. See \citet{Chiosietal1998} and \citet{Tantalo_Chiosi_2002} for detailed discussions of this issue and possible ways out. In the present models we have taken the suggestions by \citet{Chiosietal1998} and \citet{Tantalo_Chiosi_2002} into account. 

   \begin{table*}
    \begin{center}
    \caption{A few key quantities of the  model galaxies at the present time. From left to right: age in Gyr, the logarithm of the stellar mass $M_s$ in solar units,  the logarithm of the effective radius \re\ in kpc, the logarithm of the velocity dispersion $\sigma$ in km/s, the logarithm of the B luminosity $L_B$ in solar units, the logarithm of specific intensity $I_{eB}$ in $L_B/pc^2$, the logarithm of the mass to light ratio $M_s/L_B$ in solar units, $L_V$, $I_{eV}$, $M_s/L_V$ the same for the V band, the redshift of galaxy formation $z_{f}$, the asymptotic baryonic mass $M_B(t_G)$  in solar units, the infall time scale $\tau$ in Gyr, and finally the notes N where  the asterisks mean that the models take all corrections to the \citet{Fan_etal_2010} radius into account.  } 
    \label{Tab_5}
    \begin{tabular}{|r r r  r r r  r r r  r r r r r|}
     \hline  
      age   & Ms      &  Re      &$\sigma$  & $L_B$  & $I_{eB}$ &$M_s/L_B$ & $L_V$  &$I_{eV}$ &$M_s/L_V$  &  $z_{f}$ &  $M_B$ &  $\tau$&N\\
\hline   
     13.18  &  5.975  &   -0.245  &  0.410  &  4.981 &  -1.326  &  0.99  &  5.096 &  -1.211 &   0.87  &   10   &  6  & 1 & \\ 
     13.18  &  7.975  &    0.422  &  1.077  &  6.981 &  -0.659  &  0.99  &  7.102 &  -0.538 &   0.87  &   10   &  8  & 1 & \\
     13.18  &  9.975  &    1.088  &  1.744  &  8.981 &   0.008  &  0.99  &  9.096 &   0.123 &   0.87  &   10   & 10  & 1 & \\
     13.18  & 11.975  &    1.755  &  2.410  & 10.981 &   0.674  &  0.99  & 11.102 &   0.795 &   0.87  &   10   & 12  & 1 & \\
\hline     
     13.18  &   5.987 &   -0.845  &  0.716  &  4.988 &  -0.118  &  0.99  &  5.109 &   0.003 &   0.87  &   10   &  6  & 1 & *\\
     13.18  &   7.987 &   -0.179  &  1.383  &  6.988 &   0.549  &  0.99  &  7.109 &   0.670 &   0.87  &   10   &  8  & 1 & *\\
     13.18  &  9.987  &    0.488  &  2.050  &  8.988 &   1.215  &  0.99  &  9.109 &   1.337 &   0.87  &   10   & 10  & 1 & *\\
     13.18  & 11.987  &    1.155  &  2.716  & 10.988 &   1.882  &  0.99  & 11.109 &   2.003 &   0.87  &   10   & 12  & 1 & *\\
\hline     
     12.63  &  5.989  &   -0.845  &  0.717  &  5.018 &  -0.088  &  0.97  &  5.135 &   0.029 &   0.85  &    5   &  6  & 1 & *\\
     12.63  &  7.988  &   -0.178  &  1.383  &  7.011 &   0.570  &  0.97  &  7.129 &   0.688 &   0.85  &    5   &  8  & 1 & *\\
     12.63  &  9.988  &    0.489  &  2.050  &  9.011 &   1.237  &  0.97  &  9.129 &   1.354 &   0.85  &    5   & 10  & 1 & *\\
     12.63  & 11.988  &    1.155  &  2.716  & 11.011 &   1.903  &  0.97  & 11.129 &   2.021 &   0.85  &    5   & 12  & 1 & *\\
\hline    
     11.58  &  5.990  &   -0.843  &  0.716  &  5.049 &  -0.063  &  0.94  &  5.161 &   0.049 &   0.82  &    3   &  6  & 1 & *\\
     11.58  &  7.990  &   -0.176  &  1.383  &  7.049 &   0.604  &  0.94  &  7.174 &   0.750 &   0.81  &    3   &  8  & 1 & *\\
     11.58  &  9.990  &    0.491  &  2.050  &  9.049 &   1.271  &  0.94  &  9.161 &   1.382 &   0.82  &    3   & 10  & 1 & *\\
     11.58  & 11.990  &    1.157  &  2.716  & 11.049 &   1.937  &  0.94  & 11.161 &   2.049 &   0.82  &    3   & 12  & 1 & *\\        
\hline    
      7.77  &  5.989  &   -0.844  &  0.716  &  5.211 &   0.102  &  0.77  &  5.296  &  0.187 &   0.69  &    1   &  6  & 1 & *\\
      7.77  &  7.989  &   -0.341  &  1.465  &  7.211 &   1.096  &  0.77  &  7.296  &  1.181 &   0.69  &    1   &  8  & 1 & *\\
      7.77  &  9.989  &    0.489  &  2.050  &  9.211 &   1.436  &  0.77  &  9.296  &  1.521 &   0.69  &    1   & 10  & 1 & *\\
      7.77  & 11.989  &    1.156  &  2.716  & 11.211 &   2.102  &  0.77  & 11.296  &  2.187 &   0.69  &    1   & 12  & 1 & *\\
\hline     
      5.10  &  5.975  &   -0.858  &  0.716  &  5.370 &   0.289  &  0.60  &  5.427  &  0.346 &   0.54  &    0.5 &  6  & 1 & *\\
      5.10  &  7.975  &   -0.191  &  1.383  &  7.370 &   0.956  &  0.60  &  7.427  &  1.013 &   0.54  &    0.5 &  8  & 1 & *\\
      5.10  &  9.975  &    0.475  &  2.050  &  9.370 &   1.622  &  0.60  &  9.427  &  1.679 &   0.54  &    0.5 & 10  & 1 & *\\
      5.10  & 11.975  &    1.142  &  2.716  & 11.370 &   2.289  &  0.60  & 11.427  &  2.346 &   0.54  &    0.5 & 12  & 1 & *\\     
\hline      
      5.12  &  9.929  &    0.427  &  2.051  &  9.323 &   1.671  &  0.60  &  9.380  &  1.728  &  0.55  &    0.5 & 10  & 5 & *\\
\hline      
 \end{tabular}
 \end{center}
 \end{table*}
 
\begin{figure}
 \centering
 {\includegraphics[width=8.0cm, height=8.0cm]{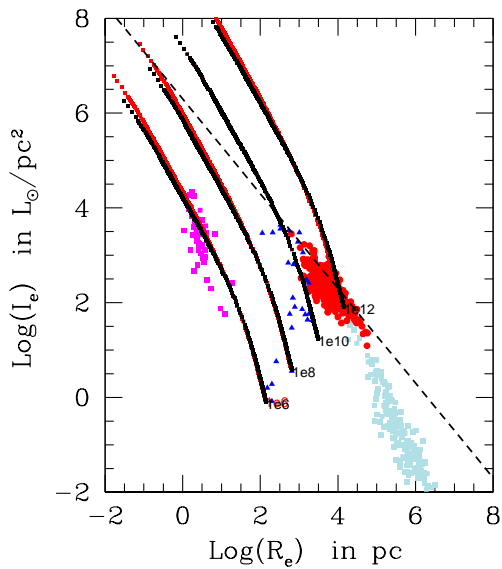} }
 \caption{ The $I_e$-$R_e$ plane  of our model galaxies with galactic winds powered by the energy input from  supernova explosions and stellar winds. There is no visible difference with respect to the same plane of models without galactic wind. The same notation, symbols and color code of Fig.\ref{fig_23} is adopted here.}
\label{fig_26} 
\end{figure}

\textsf{The role of Galactic Winds}. The main lines of the discussion for models without galactic winds holds good also for the new ones. Therefore we focus on key relations such as the \IeRe\ plane which is displayed in Fig. \ref{fig_26}. The comparison with the same plot of Fig. \ref{fig_23} shows that there is no visible difference passing from models without to those with galactic winds. The reason for that is the kind of star formation at work. Because of the short infall time scale and the dependence of $\tilde{\nu}$ on the inverse of the velocity dispersion, most of the stars are in place before the occurrence of galactic winds. To somewhat alter this trend one should change the parameter $\tau$ and make it to depend on the galaxy mass, for instance long in low mass galaxies and short in the high mass ones. To further investigate this point is beyond the aims of this study. 
 
\textsf{Role of the Initial Mass Function}. To avoid misunderstanding, we need to recall here that the present models are calculated with the classical IMF of \cite{Salpeter1955}. Therefore the $M_s/L$ ratio based on these models has this fundamental limitation and cannot by applied to investigate the problem of the FP tilt in a very general way. Our infall models can easily be adapted to include popular IMFs in literature different from the Salpeter case, see for instance  \citet[][]{Chiosietal1998}, where the IMF is let vary with the physical condition (mean density, temperature and velocity dispersion) of the gas inside a galaxy and therefore with time  for a galaxy of given mass and with time and mass in objects of different mass. However, for the aims of this study, in order to simplify this we thought it wise to rely on the classical IMF of Salpeter. If the present models were applied to the issue of the FP tilt, most likely they could account for only half of the observational tilt. This subject was specifically addressed in \citet{Chiosietal1998} with good results for the tilt of the FP of ETGs in the Virgo and Coma clusters. 

\textsf{Changing the galaxy mass and $z_{f}$}. An important feature of the models is related to the formalism in use. According to the formalism and equations widely described in \citet{Tantalo_etal_1998} all relevant physical quantities describing the model and its temporal evolution are suitably normalized to the so-called asymptotic baryonic mass $M_B(t_G)$, for instance the gas mass at time $t$ is expressed as $G_g(t) = M_g(t)/M_B(t_G)$, equally for the star mass $G_s(t) = M_s(t)/M_B(t_G)$, and the current total baryonic mass  $G_B(t) = M_B(t)/M_B(t_G)$. The amount of dark matter at any time is simply related to the current baryonic mass via the cosmic ratio (the components are intimately mixed together so that they {fall  together at the same rate}). Furthermore, the accretion rate, the star formation rate, etc. are all expressed using the same kind of normalization.  The advantage is that the time scale of mass accretion $\tau$, the cosmic ratio $f_c$, and all the rest is parameter-free, so that the only free quantity is the asymptotic baryonic mass $M_B(t_G)$. This allows us to generate models for any value of $M_B(t_G)$.

\begin{figure}
 \centering
 {\includegraphics[width=8.0cm, height=8.0cm]{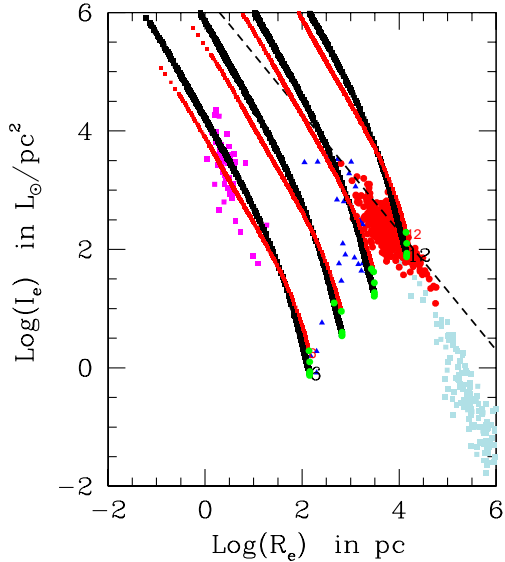} }
 \caption{ The $I_e$-$R_e$ plane of our model galaxies with different formation redshift $z_f$, namely 10, 5, 3, 1 and 0.5. The four green points of different colors are the present day stage of model galaxies whose existence began at redshifts from 0.5 to 5.  The effect is quite small.}
\label{fig_27} 
\end{figure}

All galaxy models discussed so far are calculated assuming the redshift of galaxy formation $z_{f}=10$. Other values are of course possible. Higher values are unlikely whereas lower values are much plausible. Since the age of the Universe $T_U$ depends only on the cosmological model in use and therefore is a fixed quantity, the age of a galaxy $T_G$ expressed by $T_G = T_U - T_{U}(z_{f})$ at decreasing $z_{f}$ becomes shorter. Consequently some features of the models  will change, such as total ages, radii, luminosities and specific intensities. The following values of $z_{f}$ are considered: 5, 3, 1, and 0.5 in addition to the previous set with $z_{f}=10$. The results are shown in Fig.\ref{fig_27} limited to the cases $z_{f}=5$ (black lines) and $z_{f}=0.5$ (red lines).
The case $z_{f}=10$ runs nearly over the case $z_{f}=5$. All the others are in between the case $z_{f}=5$ and $z_{f}=0.5$. From left to right, the galaxy mass is $M_B(T_G)= 10^{6} 10^{8}, 10^{10}, 10^{12} \, M_\odot$. The age increases along  each line from the top to the bottom. The final age (in Gyr) decreases from 13.19  for $z_{f}=$10  to 12.47 for $z_{f}=$5, 11.48 for $z_{f}=$3, 7.73 for $z_{f}=$1, and 5.02  for $z_{f}=$0.5. See Table \ref{Tab_5} for more information on the final stage of each model. In Fig \ref{fig_27} the final stages are represented by the green circles (some of them overlap).

From these data we derive that variations in $z_{f}$ from 10 to 0.5 yields variations in $\log(I_e)$ of about $\Delta \log (I_e) \simeq 0.5$ while the radius does not change significantly. More efficient star formation in recent times generates more luminosity and hence higher specific intensity $I_e$. This is achieved by changing $\tau$ from 1 to 5 Gyr (in the case of the $10^{10} \, M_\odot$ galaxy) yielding   $\Delta \log (I_e) \simeq 0.4$. Recent bursts of star formation either by internal causes or mergers  would also increase $I_e$. Analysing all implications of it is beyond the aims of this study. What we can say with confidence is that a significant scatter in the $I_{e}-R_{e}$ plane is likely to occur. In any case, the gross distribution of galaxies in this plane (but for GCs and GCGs) is accounted for by these models.
  
Finally, the homologous behaviour of the models and the limited effect of the formation redshift on their evolutionary behaviour make it possible to generate simulations of the distribution of large number of galaxies  in the parameter space we are investigating in practice at no cost.
 
\begin{figure}
 \centering
 {\includegraphics[width=8.0cm, height=8.0cm]{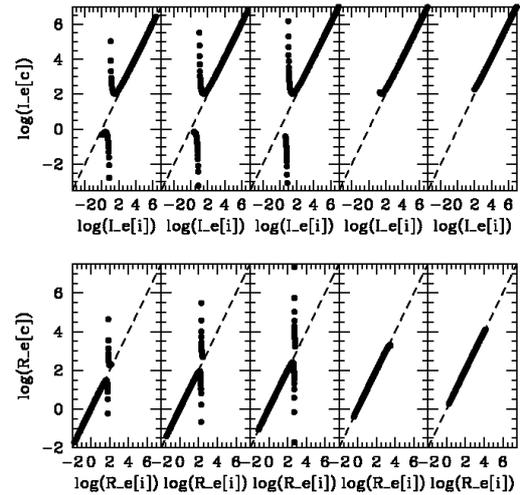} }
 \caption{ The comparison of $I_e$ and $R_e$ derived for the model galaxies (indicated by the suffix [i] and those calculated with relations (\ref{relIeSig}) and (\ref{relReSig}) for galaxies with asymptotic baryonic mass  $M_B(T_G)=$ $10^{6}, 10^7, 10^{8}, 10^{10}$, and $10^{12} \, M_\odot$ from left to right. The redshift of galaxy formation is  $z_{f}$=10.}
\label{fig_28} 
\end{figure}

\textsf{A test of consistency}. The galaxy models we have presented are based on physical assumptions such as the infall picture,  the star formation rate, the mass-radius relationship and the population synthesis governing their  luminosity in different pass-bands, that are not explicitly related to our interpretation of the parameter space of galaxies (luminosity, stellar mass and radius, velocity dispersion, and specific intensity), the FP in the multi-dimensional space  and its possible projections onto different planes that led us to the $L=L'_0 \sigma^\beta$ relationship with $L'_0 $ and $\beta$ changing from galaxy to galaxy and for each of them  also with time. On this ground we have made some detailed predictions about $L'_0 $ and $\beta$ and derived a number of equations whose solutions on one hand yield $L'_0 $ and $\beta$ as function of $L$, $M_s$, $R_e$, $\sigma$, etc. and  on the other hand allows to construct the expected relationships among pair of fundamental variables such as $I_e$ vs $R_e$, $I_e$ vs $\sigma$, $R_e$ vs $\sigma$ etc.. Among these we choose here  as an example the variables $I_e$ and $R_e$ and compare the values given by the models with those derived from  eqs. (\ref{relIeSig}) and (\ref{relReSig}). The comparison is shown in Fig. \ref{fig_28}: the top panels are for $I_e$ while the bottom panels are for $R_e$; the galaxy mass $M_B(T_G)$ is $10^{6}, 10^7, 10^{8}, 10^{10}$, and $10^{12} \, M_\odot$ from left to right (the case $M_B(T_G)=10^{7} \,\odot$ is added). On the abscissa are the input values from the models (labelled $I_e[i]$ and $R_e[i]$) and on the ordinate  the values calculated from eqs. (\ref{relIeSig}) and (\ref{relReSig}) (labelled $I_e[c]$ and  $R_e[c]$). In general there is a surprisingly good agreement between $[i]$ and $[c]$ quantities, but for some particular stages in which the $[c]$-values rapidly diverge and change sign. The cause of it resides in the analytical relationships themselves that contain various exponents (e.g. $\gamma$, $[(\beta-2) / (1+ 3/\gamma)]$, $[(\beta-2) / (3+ \gamma)]$  that in turn are functions of $\beta$ which varies in the course of evolution. In this narrow interval the disagreement is of mathematical nature with no physical implications. It simply means that these analytical relationships cannot be used with safe to derive the corresponding variables.  
 
\textsf{General remarks and preliminary conclusions}. Since the $[i]$- and $[c]$-values are nearly coincident, using the analytical relationships would predict results in the various {projection} planes we have examined identical to those obtained from using the numerical galaxy model.  The overall agreement  between  the model and analytical approach lends strong support to  the idea at the base of the analytical view, i.e. that the relation between the luminosity and velocity dispersion of a galaxy is governed by $L=L'_0 \sigma^\beta$ in which both $\beta$ and $L'_0$ vary with the galaxy mass, evolutionary stage (and hence time and redshift) and these quantities in turn are intimately related to key physical parameters such as the stellar mass and radius, the velocity dispersion (a measure of the gravitational potential well), the star formation rate, the infall time scale, and finally the ratio $M_D/M_s$. The  distribution of galaxies on the usual diagnostic planes such as FP, FJ, $I_e$-$R_e$, $M_s$-$R_e$, $L$-$R_e$, $R_e$- $\sigma$, and finally the border of the exclusion zone,  mirror the mean behaviour of galaxies each of which has its particular history and is observed in some evolutionary stage.

\begin{figure*}
 \centering
 {  \includegraphics[width=6.0cm, height=6.0cm]{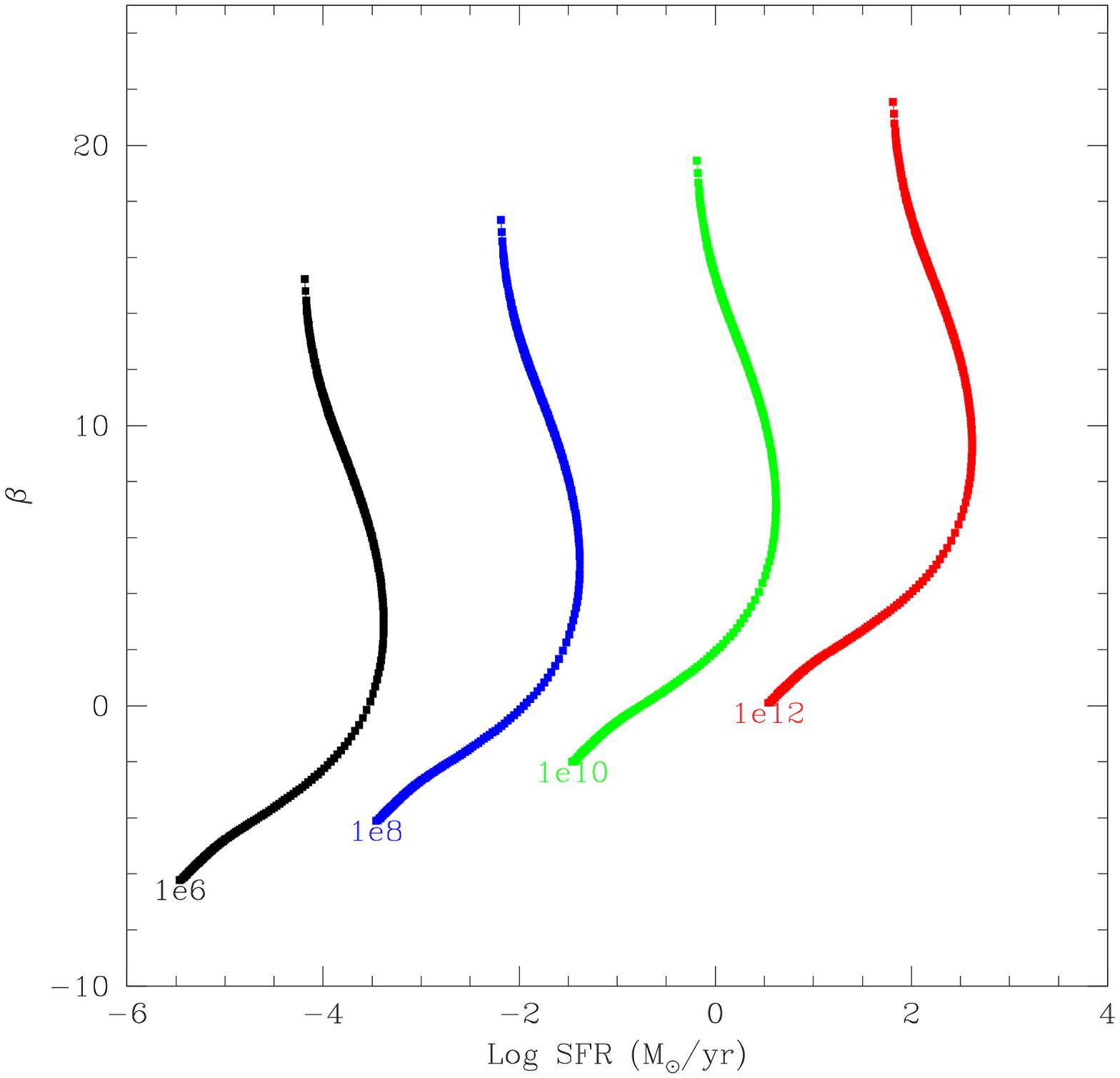} 
    \includegraphics[width=6.0cm, height=6.0cm]{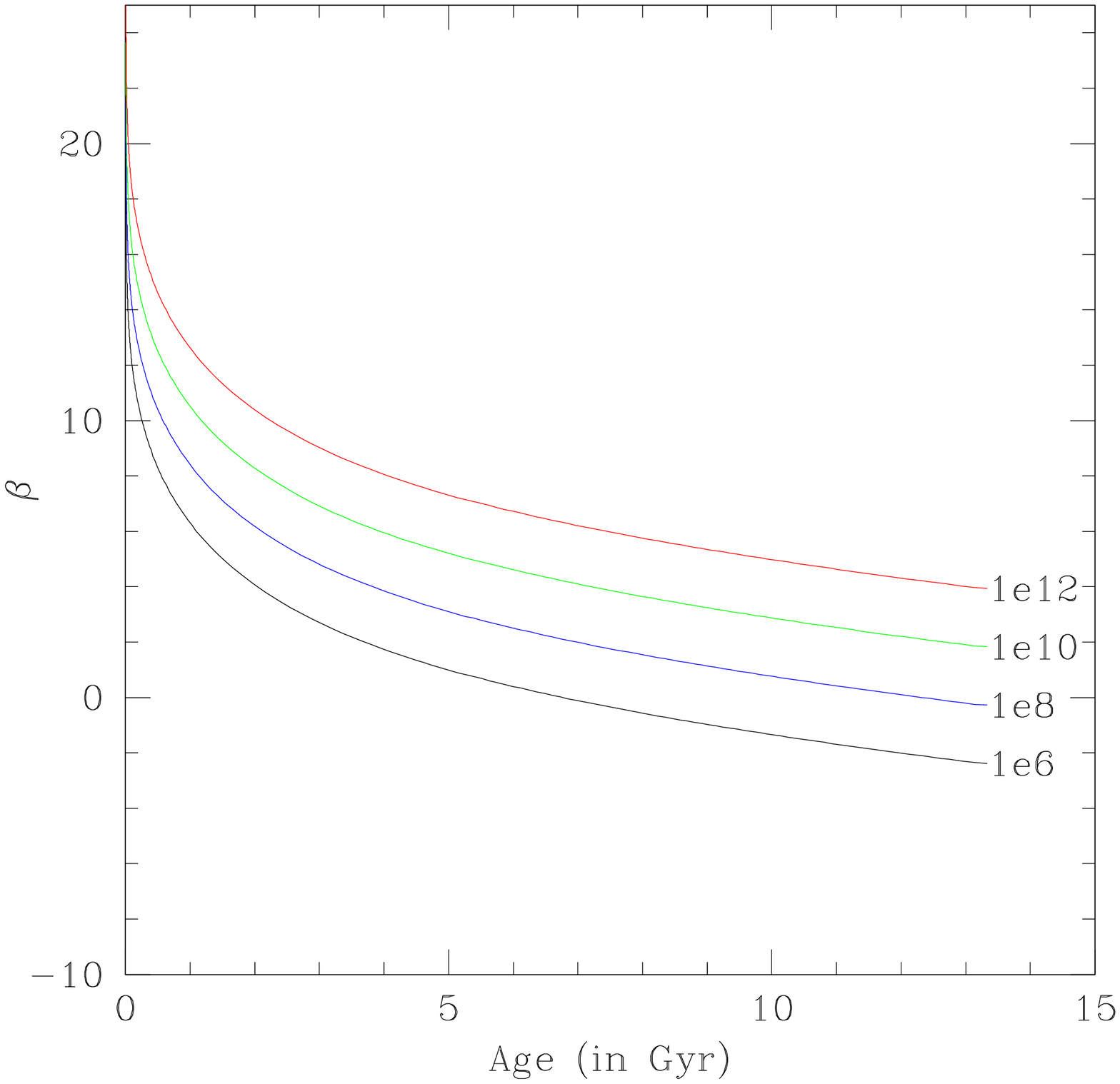}   
    \includegraphics[width=6.0cm, height=6.0cm]{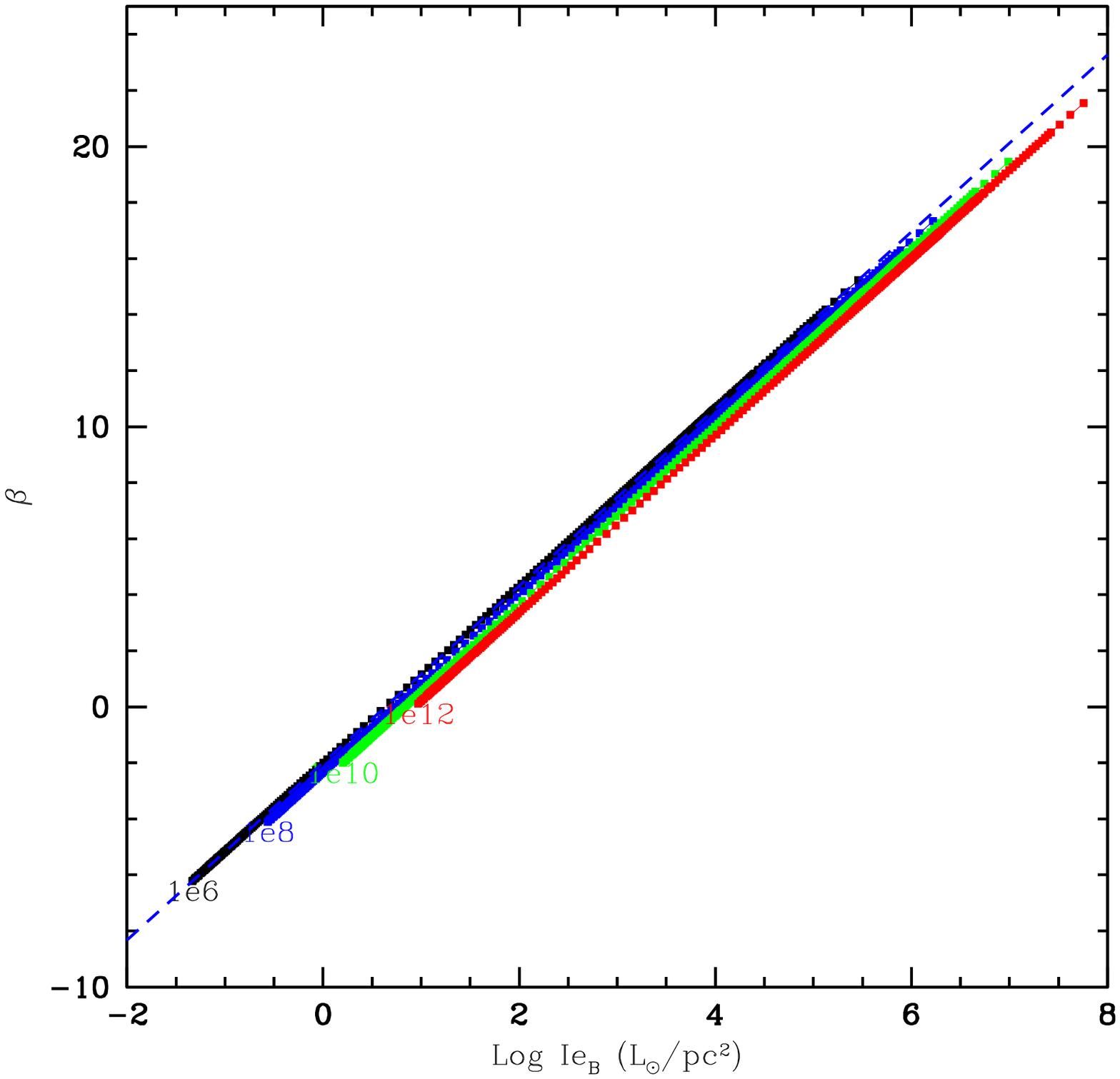} 
    }
 \caption{Left Panel: The relationship between $\beta$ and the star formation rate (SFR). Each  curve labelled by  $M_{B}(T_{G})$, is identified by a different colour according to the color-code adopted in previous figures. The total baryonic mass is the homology parameter separating each curve from the others. Much similar trends are found for the luminosity $L_B$ and $L_V$, and the velocity  dispersion $\sigma$ that are not displayed here for the sake of brevity. In all three relations  $\beta$ mirrors the behavior of the SFR, the luminosity in turn, and finally the velocity dispersion.  The SFR is in $M_\odot/yr$. Middle Panel:  The relationship between $\beta$ and age (in Gyr). Symbols and color-code have the same meaning as in the left panel. Right Panel: the relationship between $\beta$ and  $I_{eB}$  in $L_\odot /pc^2$. The lines corresponding to different masses of galaxies have been displayed by shifting each of them by 0.1; in reality they collapse to a  unique curve given  by $\beta = 3.159 Log(I_{eB}) - 2.003$. The long dashed line is the best fit of the theoretical data. Identical relation can be found for $I_{eV}$: same slope but slightly different zero point.}
\label{fig_29} 
\end{figure*}

\begin{figure}
 \centering
 {\includegraphics[width=8.0cm, height=8.0cm]{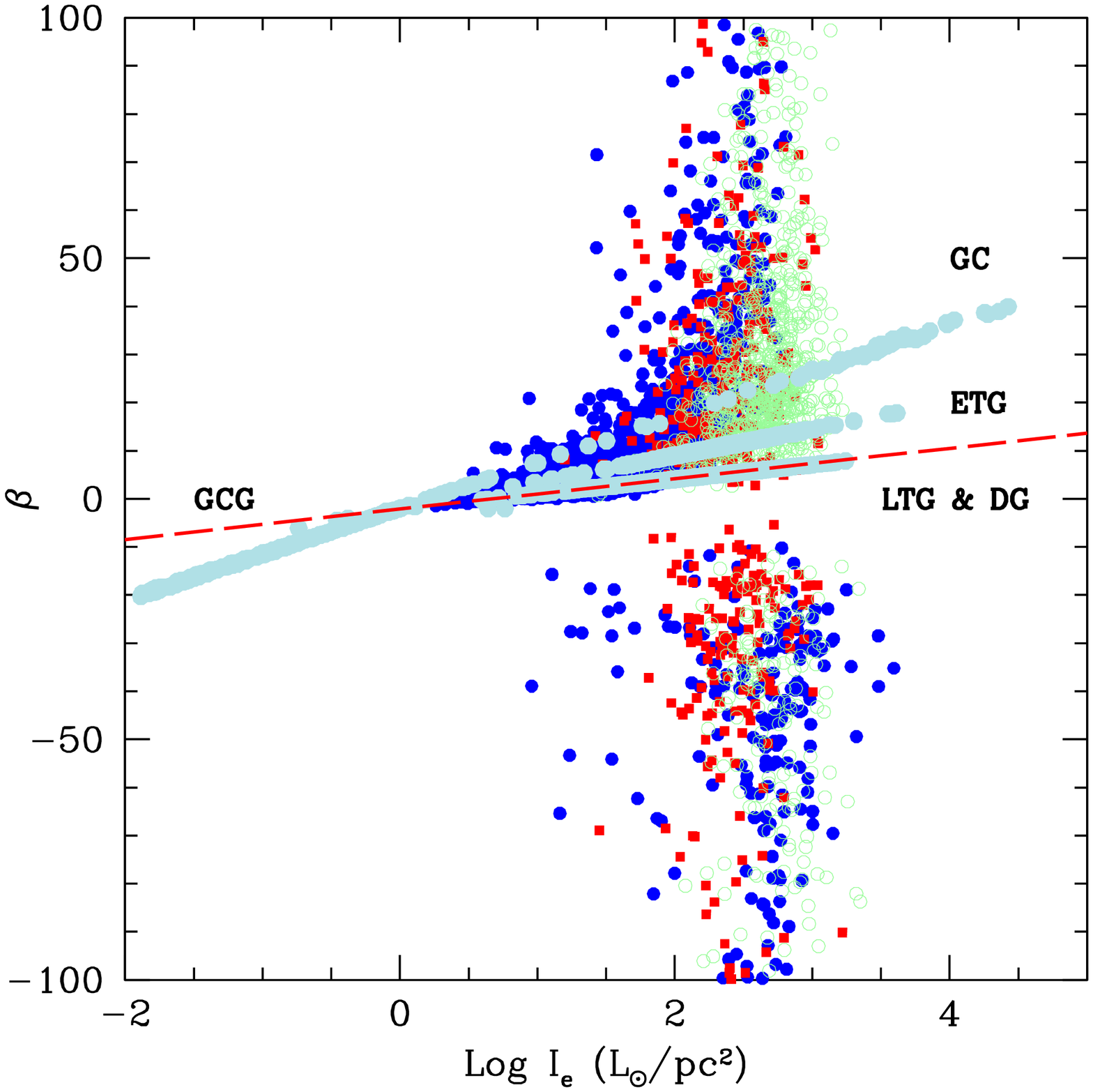}}
 \caption{ 
 The $\beta$ - $I_{eV}$ plane: data and theoretical models. Data from the different sources are plotted: (i) The  powder-blue points from \citet{Burstein_etal_1997}; three sequences are seen: the sequence of GCGs \& GCs, the one of ETGs (no evidence of star formation), and the one of LTGs and DGs (evidence of ongoing star formation). By construction, the data of  \citet{Burstein_etal_1997} are well behaved with no evidence of dispersion. (ii) The red squares are the WINGS data showing large dispersion in both coordinates, $\log I_e$ is always positive, and $\beta$ can be very large both positive and negative. (iii) the open green circles are the ETGs of \citet{Bernardi_etal_2010} however limited to $z \leq 0.02$.
The Illustris-1 model galaxies are indicated by the blue dots; their distribution closely mimics that of the observational data. Finally, the long dashed red line shows the present-day position on the $\beta$-$I_e$ plane of our models for the reference case (with $\tau=1$ Gyr, $z_f=10$ and no galactic winds). This line coincides with the lower border of the Illustris-1 distribution in the $\beta>0$ hemi-plane.
}
\label{fig_30} 
\end{figure}

\section{The important role of $\beta$}

In this section we cast light on the role of $\beta$. To this aim we adopt the reference case ($z_f=10$ and $\tau=1$) and leave the issue of galactic winds aside. For this case we present a few basic relationships among $\beta$ and other important parameters namely the SFR (in $M_\odot/yr$),  the age (in Gyr), and  the specific intensity $I_{eB}$ or $I_{eV}$ (in $L_\odot/pc^2$). These relationships are shown in Fig. \ref{fig_29}.  In the left panel, the homologous nature of the galaxy models is evident: all curves have  the same shape, but each one is separated from all the others  by the homology parameter, namely the total baryonic mass at the present age $M_{B}(T_{G})$ annotated along each curve.  The temporal evolution occurs from the top to the bottom of each curve.  Identical behaviors are found between $\beta$ and the luminosity $L_B$ or $L_V$ (in $L_\odot$), and the velocity dispersion $\sigma$ (in km/s). However, these relationships are not shown here for the sake of brevity. The central panel of Fig. \ref{fig_29}, showing the variation of $\beta$ with the age, still displays the dependence of the results on the homology parameter and thus there are four different curves one for each value of the $M_{B}(T_{G})$. 
Finally, in the right panel we show the dependence of $\beta$ on the  surface brightness $I_{eB}$; all curves collapse to a single relation, the homology is destroyed by the underlying relationship between the mass and the effective radius of the models. A similar relation is found between between $\beta$  and $I_{eV}$. The analytical relations between $\beta$ and $I_{e}$ are given by 

\begin{eqnarray}
 \beta &=& 3.159 Log(I_{eB}) - 2.003  \\
 \beta &=& 3.159 Log(I_{eV}) - 1.900. 
 \label{beta_Ie}
\end{eqnarray}

\noindent
The evolution along each line is from top-right to bottom-left and the present stage is the last point where $M_{B}(T_{G})$ is annotated.
The above relations indicate both the path followed by a single galaxy in the course of its history and also  the locus on the $\beta$-$I_e$ plane of galaxies of different mass observed at the present age. There is no appreciable effect of different formation redshifts, at least in the interval $10 \geq z_f \geq 1$ nor of different accretion timescale $\tau$ in the interval  $1 \leq \tau \leq 5$ Gyr. We estimate a total effect on $\beta$ by redshift $z_f$ and accretion time scale $\tau$ of the order $\Delta\beta \simeq 2$ over the interval of $I_{eV}$ of interest here. 

The linear relation between $\beta$ and $I_e$ shown by our models is a very intriguing result that demands a thorough analysis because observational data and numerical hierarchical models seem to indicate a different picture.  The situation is best illustrated by Fig. \ref{fig_30} comparing data and models from different sources. On the observational side we have three data-sets: \citet{Burstein_etal_1997}, WINGS, and \citet{Bernardi_etal_2010}. The last two (mainly devoted to ETGs) are based on equivalent methods to estimate $R_e$, and therefore yield similar results as far as the  $\beta$-$I_e$ plane is concerned. In contrast, the first one  that contains objects going from GCs to DGs, LTGs, ETGs and GCGs, differs in the method used to derive the effective radius, and consequently yields different relationships in the $\beta$-$I_e$ plane. Owing to  this, some preliminary remarks are needed. First of all, the  data of \citet{Burstein_etal_1997} are in the B-band so that must be transformed into the V-band. This is made by means of the relation $$ \log L_V = 0.4 [(B-V)_0 -0.65] + \log L_B,$$ where the luminosities are in solar units, $(B-V)_0$ is the colour, and -0.65 is the difference between the B and V photometric constants (5.48 and 4.83 respectively). Second, recalling that the luminosity $L_e$  given by \citet{Burstein_etal_1997} is the amount of light falling within the effective radius $R_e$,  where half the total luminosity is found, we scale it by a factor of two  to  make it consistent with the definition of $I_e$ we have adopted. 

The observational data for $Re$, $L_V$,  $M_s$, ($M_s/L_V$), $\sigma$, and $I_{eV}$ are fed to the system of equations (\ref{eqsyst})  and the solutions $\beta$ and $L'_0$ are derived.
The  powder-blue points n Fig. \ref{fig_30} are the \citet{Burstein_etal_1997} data; three sequences are seen: the GCG-GC sequence,  the one of ETGs  (no evidence of star formation), and the one of LTGs and DGs (evidence of ongoing star formation). By construction, the data of \citet{Burstein_etal_1997} are well behaved with no evidence of dispersion. The red squares are the WINGS data showing large dispersion in both coordinates, $\log I_e$ is always positive, and $\beta$ can be very large both positive and negative. Much similar results are found with the \citet{Bernardi_etal_2010} data, the open green circles.
The Illustris-1 model galaxies are indicated by the blue dots; their distribution closely mimics that of the observational data. Finally, the long dashed red line shows the present-day position on the $\beta$-$I_e$ plane of our models for the reference case (with $\tau=1$ Gyr, $z_f=10$ and no galactic winds). This line coincides with the lower border of the Illustris-1 distribution in the $\beta>0$ hemi-plane.  Choosing different values of $z_f$ in the interval $0.5 \leq z_f \leq 10$ does not shift significantly the line predicted at the present time. Equally for the effect of galactic winds. Lumping all these effects together we expect a typical width of this border line of about $\Delta \beta \simeq 10$ over the $I_{eV}$ interval of interest here.  

From this preliminary comparison we may conclude that mutual consistency among different sources of data and of these latter with models exists.

The major issue now is to understand the physical causes of the large dispersion in $\beta$ for all values of $I_e \geq 1$. Looking at the eqs. (\ref{eqbeta}), providing the solutions $\beta$ and $log(L'_0)$ of our equations, one notes that  under suitable conditions  the term $1- 2A'/A$ at the denominator of $log(L'_0)$ gets very close to zero, consequently  $\beta$  can be either very large and positive or large and negative. As already discussed in Sect. \ref{sec:3} when this happens the system is in conditions of strict virialization. The sign of $\beta$ depends on the particular history of the constituent variables ($M_s$, $R_e$, $L$, and  $I_e$), in other words whether the term  $2A'/A$ is tending to 1 from below ($\beta > 0)$ or above 1 ($\beta<0$). From an operative point of view we may define "state close to strict virialization" when $|\beta| > 20$. This would account for the gap on the negative hemi-plane of Fig. \ref{fig_30}.

Do data and models ever reach the condition of full virialization indicated $\beta \Rightarrow \pm \infty$ or do they remain somewhat far it? The answer is that both possibilities occur.

On the observational side, given any galaxy for which the set of parameters ($M_s$, $R_e$, $L$, $I_e$ and $\sigma$) has been measured, it is not granted that they would satisfy the virialization condition. The major uncertainties are with $M_s$ and $R_e$ and $I_e$ in turn. Therefore, many of them crowd in the interval $-1 \leq \beta \leq 20$ which implies deviations from virial equilibrium. However with the present data, it is not possible to say whether this is due to insufficient accuracy in the parameters determinations or real deviations from virial equilibrium due to recent mergers, harassment, loss of mass, interactions etc. However, many other galaxies in both hemi-planes with $|\beta| > 20$ which is a strong indication that they are close to virial equilibrium. 

On the theoretical side, our model galaxies with infall (no dynamics in them) seem to be in a state far from strict virialization. This is suggested by the small values of $\beta$ reached at the present time. The reason for that resides in the way the models are built up. In brief, mass point description with no dynamics is adopted, the total mass is assigned (via the accretion law), the stellar mass is derived from star formation, the effective radius is estimated from a suitable relationship, the luminosity is evaluated  from the stellar mass and a mean  luminosity-age relationship for a fictitious SSP with mean metal content (the difference with respect to the luminosity correctly derived from the theory of population synthesis  via the history of star formation and metal enrichment is not large but still significant), finally the velocity dispersion is derived from the VT with the current values of $M_s$ and $R_e$. The major uncertainties are in $R_e$ and $L$. Therefore our set of basic parameters not necessarily can fulfill all the requirements  imposed by the VT. In consequence, our $\beta$s are always quite small (say smaller than 25-30) implying that full virialization is not reached. However, this failure is not as severe as it appears  because small adjustments of  $R_e$ and $L$ are possible while the models are successful in many other aspects.

The situation is much better with the Illustris-1 models where if a good number of galaxies have $\beta < 25-30$ as in the case of our models,  still a large number of objects is clearly seen in regime of strict virialization because of their high positive and/or negative $\beta$s. The inclusion of real dynamics and the hierarchical scenario at work provide much better conditions to bring the action of virialization into evidence. In the  hierarchical scenario, mergers, ablation of stars and gas, {harassment}, secondary star formation, inflation of dimension by energy injections of various kinds, etc. induce strong variations on the structural parameters and hence strong temporary deviations from the virial conditions. However, once {this  happened}, the viral conditions  can be soon recovered over a suitable timescale. This can be  short or long  depending on the amount of mass engaged in the secondary star forming activity and the amount of time elapsed since the star forming event took place \citep[see the burst experiments in][]{Chiosi_Carraro_2002, Tantalo_Chiosi_2004}.  As  a consequence of all this,  detecting systems on their way back to virial equilibrium is likely a frequent event thus explaining the high dispersion seen on the $\beta$-$I_e$ plane.  

In principle, the value of $\beta$ evaluated for each galaxy could provide a useful hint about the equilibrium state reached by the system. Most likely, the condition of strict virial equilibrium is a transient phenomenon that could occur several times during the life of a galaxy. This is perhaps suggested by the high numbers of galaxies with both low and positive values of $\beta$ and high positive/negative values  of $\beta$.

\section{Discussion and conclusions}\label{sec:5}

The aim of this paper is to prove that the difficulties encountered in understanding the distribution of galaxies on the FP in the parameter space $\sigma$, $R_e$ $I_e$ and its projections on the three coordinate planes, can be removed by introducing  the \Lsigb\ relation as a proxy of evolution, in which $\beta$ and $L'_0$ vary from galaxy to galaxy and for each of them in course of time  
\citep[see][for previous efforts along this line of thought]{donofrioetal17,Donofrioetal2019,Donofrioetal2020,DonofrioChiosi2021}.The continuous variation of $\beta$ and $L'_0$  traces {the  path} followed by each ETG in the \Lsig\ plane. The  \Lsigb\  law together  with the VT yield a set of relations $R_e$-$\sigma$, $I_e$-$\sigma$ and $R_e$-$M_s$ that nicely reproduce the data and suggest the existence of a system of two equations in the unknowns $\beta$ and $L'_0$  with coefficients functions of $M_s$, $R_e$, $L$, and $I_e$ that for each galaxy determine the value of $\beta$ and $L'_0$. With {the aid} of these relations we can  determine the instantaneous position and direction of  a galaxy on the FP and  the projection planes. 

The analysis is made in two steps. In the first one,  the problem is addressed from an observational point of view, inferring from the data the expected position and evolutionary direction of a galaxy in the various planes and owing to the large number of galaxies in the samples the  range of values spanned by $\beta$ and $L'_0$ is determined. In the second step, simple models of galaxy formation, structure and evolution are set up, the basic equations are solved at each time step of a galaxy's lifetime  so that the history of  $\beta$ and $L'_0$  is known. Basing on these results, the various projection planes are examined finding consistency between observational data and theoretical models.  The same procedure is applied to literature galaxy models calculated in the framework of  the hierarchical scenario. The theoretical results are compared with the observational data and good mutual agreement is found. Basing on this, we conclude that the starting hypothesis about the real existence of the \Lsigb\ relation is correct.  In more detail, the present analysis has clarified the following issues:

\begin{enumerate}
\item The FP can be understood as the average of the single FP-like relations valid for each galaxy (eq. \ref{eqfege}). The coefficients of the FP-like relation are function of $\beta$. This means that the FP must evolve with redshift and that its coefficients depend on the adopted waveband (in which observations are taken) and on the nature of the data sample (how many ETGs are included) as confirmed by the current observational data;
    
\item All the features of the FP projections can be explained at the same time. This includes: 1) the curvature of the relations, that turns out to depend on the existence of positive and negative values of $\beta$, and 2), the existence of the ZoE, \ie\ the line marking the separation between the permitted and forbidden  regions in these planes. No galaxies can reside in the ZoE. 

\item The FP and all its projections, such as for instance the classical Faber-Jackson relation, are instantaneous pictures of the present-day situation. They should change with redshift hence lifetime of galaxies. 

\item The ZoE is obtained in a natural way as the only possible  evolutionary path for objects with large positive and negative $\beta$'s that are well virialized. These objects in general have stopped their star formation long ago and their luminosity progressively decreases. When ETGs become passive quenched objects, with a luminosity decreasing at nearly constant $\sigma$, the galaxies can only move in one direction, that given by the large positive and negative values of $\beta$.

\item The infall model galaxies built here, although lacking the dynamical component,  are in reasonable agreement with observations and support the idea that $\beta$ and $L'_0$ vary across time, and therefore that the \Lsigb\ law is a plausible empirical relation accounting of the variation occurred in a galaxy.

\item All the diagrams built using the structural parameters are sensitive to the temporal evolution of galaxies, simply because each individual object moves in a different way according to the value of $\beta$. 

\item Both observations and theory suggest that the \Lsigb\ relation provides an empirical way of capturing  the temporal evolution of ETGs (and probably of late-type objects) because  the values of $\beta$ are related to the history of mass assembly and luminosity evolution. Because of it we are tempted to suggest that eqs. (\ref{eqsyst}) are two important equations governing the evolution of ETGs.

\item Finally, the large negative and positive values of $\beta$ of some galaxies can be considered as the signature that these system are very close to the virial equilibrium, i.e. their basic parameters $R_e$, $L_{\Delta\lambda}$,  $M_s$, ($M_s/L_{\Delta\lambda}$), $\sigma$, and $I_{e,\Delta\lambda}$ are such that the strict virial condition is verified. The situation is likely transient because both internal and/or external events may alter one or more parameter so that the strict virial condition is no longer verified. Since the recovery time can vary a lot from galaxy to galaxy, this ideal situation has an ample range of occurrence probabilities, from frequent in some galaxies to never in others. The parameter $\beta$ can be taken as the signature of how far is { the system} from full virialization.   
\end{enumerate}

\begin{acknowledgements}
      The authors thank the anonymous referee for his/her suggestions and comments.
\end{acknowledgements}

   \bibliographystyle{aa} 
   \bibliography{FP.bib} 
 
\end{document}